\documentclass[11pt, a4paper]{article}
\usepackage[labelfont=bf,textfont=normal]{caption}
 \usepackage[font=small, format=hang, margin={1cm, 1cm}]{caption}
\usepackage[a4paper, left=3cm,right=3cm]{geometry}
\usepackage{amsfonts} 
\usepackage{amsmath} 
\usepackage{mathrsfs}  %
\usepackage{amssymb}
\usepackage{mathtools}
\usepackage{relsize}
\usepackage{setspace}   
\usepackage{color}  
\usepackage{wasysym}
\usepackage{booktabs}
\usepackage[utf8,applemac]{inputenc} 
\usepackage{tensor}
\usepackage{cite}
\usepackage{tikz}
\usetikzlibrary{calc, arrows.meta}
\usepackage{graphicx}% Include figure files
\usepackage{subfig}
\usepackage{caption}
\usepackage[normalem]{ulem}
\usepackage{hyperref}
\usepackage{epstopdf}
\usepackage{comment}

\numberwithin{equation}{section}
\usepackage{dcolumn}% Align table columns on decimal point
\usepackage{bm}% bold math
\usepackage[toc,page]{appendix}

\def\ndelta{\delta\hspace{-0.50em}\slash\hspace{-0.05em} }

\hypersetup{
    colorlinks,%
    citecolor=blue,%
    filecolor=blue,%
    linkcolor=blue,%
   urlcolor=blue,
   linktoc=page
}

\begin{document}

%\maketitle
%\abstract{[...]}

\setcounter{tocdepth}{2}

\begin{titlepage}

%\begin{flushright}\vspace{-3cm}
%{\small
%%{\tt arXiv:yymm.nnnn} \\
%\today }\end{flushright}
%\vspace{0.5cm}
$ $
\vspace{20pt}

\begin{center}
%{ \LARGE{\bf{Asymptotic Symmetries in Gauge Theories \\ and the BMS Group}}}
{ \LARGE{\bf{Asymptotic Symmetries in the Gauge Fixing Approach and the BMS Group}}} 

 \vspace{5mm}

\vspace{2mm} 
\centerline{\large{\bf{Romain Ruzziconi\footnote{e-mail: rruzzico@ulb.ac.be}}}}

\vspace{2mm}
\normalsize
\bigskip\medskip
\textit{Universit\'{e} Libre de Bruxelles and International Solvay Institutes\\
CP 231, B-1050 Brussels, Belgium\\
\vspace{2mm}
}
\vspace{25mm}
%\vfil
%\pacs{04.70.Dy}

\begin{abstract}

These notes are an introduction to asymptotic symmetries in gauge theories, with a focus on general relativity in four dimensions. We explain how to impose consistent sets of boundary conditions in the gauge fixing approach and how to derive the asymptotic symmetry parameters. The different procedures to obtain the associated charges are presented. As an illustration of these general concepts, the examples of four-dimensional general relativity in asymptotically (locally) (A)dS$_4$ and asymptotically flat spacetimes are covered. This enables us to discuss the different extensions of the Bondi-Metzner-Sachs-van der Burg (BMS) group and their relevance for holography, soft gravitons theorems, memory effects, and black hole information paradox. These notes are based on lectures given at the \textit{XV Modave Summer School in Mathematical Physics}. 

\end{abstract}

%\pacs{04.65.+e,04.70.-s,11.30.-j,12.10.-g}

\end{center}
%%%%%%%%%%%%%%%%%%%%%%%%%%%%%%%%%%%%%%%%%%%%%%%%%%%%%%%%%%%%%%%%%%%%%%%%%%%%%%%%%%%%%%%%

\end{titlepage}

\newpage
\tableofcontents

\newpage
\section{Introduction}

Gauge theories are of major importance in physics since they are involved in the fundamental description of nature. The Standard Model of particle physics and the theory of general relativity are two examples of gauge theories offering never-equaled observational predictions for our universe. Furthermore, gauge theories provide a mathematical framework that enables us to understand the deepest foundations of our physical theories.

The study of asymptotic symmetries in gauge theories is an old subject that has recently known renewed interest. A first direction is motivated by the AdS/CFT correspondence where the asymptotic symmetries of the gravity theory in the bulk spacetime correspond to the global symmetries of the dual quantum field theory through the holographic dictionary \cite{tHooft:1993dmi , Susskind:1994vu , Maldacena:1997re , Witten:1998qj , Aharony:1999ti}. A strong control on asymptotic symmetries allows us to investigate new interesting holographic dualities. A second direction is driven by the recently-established connections between asymptotic symmetries, soft theorems and memory effects \cite{Strominger:2017zoo}. These connections furnish crucial information about the infrared structure of quantized gauge theories. In gravity, they may be relevant to solve the long-standing problem of black hole information paradox \cite{Hawking:1976ra , Hawking:2016msc , Hawking:2016sgy , Haco:2018ske , Haco:2019ggi}. 

%It has long been known that theories exhibiting some global symmetries have associated conserved charges related through the Noether procedure \cite{Noether1918}. However, the problem of constructing systematically meaningful charges related to gauge symmetries, such as the electric charge in electrodynamics or energy and momentum in general relativity, has long been an open question. The resolution of this long-standing problem is intimately related to the study of asymptotic symmetries \cite{Regge:1974zd , Barnich:1994db , Barnich:1995ap , Barnich:2001jy}. 
%
%Two important directions of development have recently been noted on the topic of asymptotic symmetries in gauge theories. The first is motivated by the holographic principle and the AdS/CFT correspondence \cite{tHooft:1993dmi , Susskind:1994vu , Maldacena:1997re}. The second is driven by the connections between asymptotic symmetries, soft theorems, and memory effects \cite{Strominger:2017zoo}.  

Several approaches exist regarding asymptotic symmetries in gauge theories and the construction of associated charges. The aim of these notes is to provide a self-consistent introduction on how to impose boundary conditions in a generic gauge theory, derive the asymptotic symmetry algebra, and compute the associated surface charges. We discuss these points in the gauge fixing approach. Indeed, despite this approach being widely used in the literature, there are few references discussing the complete procedure for a general gauge theory. To illustrate the abstract definitions and relevant results, we discuss in detail the examples of general relativity in asymptotically (locally) (A)dS$_4$ and asymptotically flat spacetimes. These examples are interesting because they involve all the subtleties of the procedure. The notes aim to be pedagogical and are based on lectures given at the XV Modave Summer School in Mathematical Physics. 

In section \ref{Definitions of asymptotics}, we briefly mention the different main frameworks to study asymptotic symmetries in gauge theories. Thereafter, in section \ref{Asymptotic symmetries in the gauge fixing approach}, we focus on the gauge fixing approach. We explain the conditions under which a given gauge fixing is suitable to study asymptotic symmetries. Then, we discuss how to impose consistent boundary conditions, the associated solution space, and how to derive the asymptotic symmetry algebra. In section \ref{Surface charges}, after some digressions through the Noether procedure to construct charges associated with global symmetries, we explain what the analogue of this construction for gauge symmetries is. In particular, the Barnich-Brandt prescription is discussed and related to the covariant phase space methods in the context of diffeomorphism-invariant theories. In section \ref{Applications}, we review some recent applications of asymptotic symmetries in the context of holography and the infrared sector of gauge theories. Finally, these notes are accompanied by two appendices. Appendix \ref{Diffeomorphism between Bondi and Fefferman-Graham gauges} is a quick summary of the matching between Bondi and Fefferman-Graham gauges in general relativity. Appendix \ref{Useful results} contains some important definitions and conventions about the jet bundles and homotopy operators widely used in the text. 
   
Many reviews related to asymptotic symmetries complementary to these notes exist in the literature: see, for example, \cite{Newman:1981fn, Compere:2019qed , Strominger:2017zoo , Barnich:2010xq, Barnich:2018gdh ,  Riegler:2017fqv , Oblak:2016eij ,  Penrose:1986ca}.     

\section{Definitions of asymptotics}
\label{Definitions of asymptotics}

Several frameworks exist to impose boundary conditions in gauge theories. Some of them are mentioned next. 

\subsection{Geometric approach}

The geometric approach of boundary conditions was initiated by Penrose, who introduced the techniques of conformal compactification to study general relativity in asymptotically flat spacetimes at null infinity \cite{Penrose:1962ij, Penrose:1964ge}. According to this perspective, the boundary conditions are defined by requiring that certain data on a fixed boundary be preserved. The asymptotic symmetry group $G$ is then defined as the quotient:
\begin{equation}
G = \frac{\text{Gauge transformation preserving the boundary conditions}}{\text{Trivial gauge transformations}} ,
\label{ASG def 1}
\end{equation} where the trivial gauge transformations are the gauge transformations that reduce to the identity on the boundary. In other words, the asymptotic symmetry group is isomorphic to the group of gauge transformations induced on the boundary which preserve the given data. This is the \textit{weak} definition of the asymptotic symmetry group. A \textit{stronger} definition of the asymptotic symmetry group is given by the quotient \eqref{ASG def 1}, where the trivial gauge transformations are now the gauge transformations that have associated vanishing charges. 

The geometric approach was essentially used in gravity theory and led to much progress in the study of symmetries and symplectic structures for asymptotically flat spacetimes at null infinity \cite{Hansen:1978jz, Ashtekar:2014zsa, Ashtekar:1981bq , Dray:1984rfa , Herfray:2020rvq} and spatial infinity \cite{Ashtekar:1978zz , Ashtekar:1991vb}. It was also considered to study asymptotically (A)dS spacetimes \cite{Ashtekar:2014zfa, Ashtekar:2015lla ,Ashtekar:2015lxa  , Ashtekar:2019khv }. Moreover, this framework was recently applied to study boundary conditions and associated phase spaces on null hypersurfaces \cite{Chandrasekaran:2018aop}. 

The advantage of this approach is that it is manifestly gauge invariant, since we do not refer to any particular coordinate system to impose the boundary conditions. Furthermore, the geometric interpretation of the symmetries is transparent. The weak point is that the definition of boundary conditions is rigid. It is a non-trivial task to modify a given set of boundary conditions in this framework to highlight new asymptotic symmetries. It is often \textit{a posteriori} that boundary conditions are defined in this framework, after having obtained the results in coordinates. 

\subsection{Gauge fixing approach}

A gauge theory has redundant degrees of freedom. The gauge fixing approach consists in using the gauge freedom of the theory to impose some constraints on the fields. This enables one to quotient the field space to eliminate some of the unphysical or pure gauge redundancies in the theory. For a given gauge theory, an \textit{appropriate gauge fixing} (where appropriate will be defined below) still allows some redundancy. For example, in electrodynamics, the gauge field $A_\mu$ transforms as $A_\mu \to A_\mu + \partial_\mu \alpha$ ($\alpha$ is a function of the spacetime coordinates) under a gauge transformation. The Lorenz gauge is defined by setting $\partial_\mu A^\mu = 0$. This gauge can always be reached using the gauge redundancy, since $\partial_\mu \partial^\mu \alpha = - \partial_\nu A^\nu$ always admits a solution for $\alpha$, regardless of the exact form of $A_\mu$. However, residual gauge transformations remain that preserve the Lorenz gauge. These are given by $A_\mu \to A_\mu + \partial_\mu \beta$, where $\beta$ is a function of the spacetime coordinates satisfying $\partial_\mu \partial^\mu \beta = 0$ (see, e.g., \cite{Jackson:1998nia}). The same phenomenon occurs in general relativity where spacetime diffeomorphisms can be performed to reach a particular gauge defined by some conditions imposed on the metric $g_{\mu\nu}$. Some explicit examples are discussed below. 

Then, the boundary conditions are imposed on the fields of the theory written in the chosen gauge. The \textit{weak} version of the definition of the asymptotic symmetry group is given by 
\begin{equation}
G_{\text{weak}} = 
\left[
\begin{array}{l}
\text{Residual gauge diffeomorphisms} \\
\text{preserving the boundary conditions.}
\end{array}
\right]
\label{ASG def 2}
\end{equation} Intuitively, the gauge fixing procedure eliminates part of the pure gauge degrees of freedom, namely, the trivial gauge transformations defined under \eqref{ASG def 1}. Therefore, fixing the gauge is similar to taking the quotient as in equation \eqref{ASG def 1}, and the two definitions of asymptotic symmetry groups coincide in most of the practical situations. As in the geometric approach, a stronger version of the asymptotic symmetry group exists and is given by 
%\begin{equation}
%G_{\text{strong}} = \text{Residual gauge diffeomorphisms preserving the boundary conditions} \\
%\text{with associated non-vanishing charges}
%\label{ASG def 3}
%\end{equation}
\begin{equation}
G_{\text{strong}} = 
\left[
\begin{array}{l}
\text{Residual gauge diffeomorphisms preserving the boundary} \\ 
\text{conditions with associated non-vanishing charges.}
\end{array} 
\right]
\label{ASG def 3}
\end{equation} Notice that $G_{\text{strong}} \subseteq G_{\text{weak}}$\footnote{One of the most striking examples of the difference between the weak and the strong definitions of the asymptotic symmetry group is given by considering Neumann boundary conditions in asymptotically AdS$_{d+1}$ spacetimes. Indeed, in this situation, we have $G_{\text{weak}} = \text{Diff($\mathbb{R} \times S^{d-1}$)}$, and $G_{\text{strong}}$ is trivial \cite{Compere:2008us}.}.

The advantage of the gauge fixing approach is that it is highly flexible to impose boundary conditions, since we are working with explicit expressions in coordinates. For example, the BMS group in four dimensions was first discovered in this framework \cite{Bondi:1962px , Sachs:1962wk , Sachs:1962zza}. Furthermore, a gauge fixing is a local consideration (i.e. it holds in a coordinate patch of the spacetime). Therefore, the global considerations related to the topology are not directly relevant in this analysis, thereby allowing further flexibility. For example, as we will discuss in subsection \ref{Asymptotic symmetry algebra}, this allowed to consider singular extensions of the BMS group: the Virasoro $\times$ Virasoro superrotations \cite{Barnich:2009se, Barnich:2011ct}. These new asymptotic symmetries are well-defined locally; however, they have poles on the celestial sphere. In the geometric approach, one would have to modify the topology of the spacetime boundary to allow these superrotations by considering some punctured celestial sphere \cite{Strominger:2016wns , Barnich:2017ubf}. The weakness of this approach is that it is not manifestly gauge invariant. Hence, even if the gauge fixing approach is often preferred to unveil new boundary conditions and symmetries, the geometric approach is complementary and necessary to make the gauge invariance of the results manifest. In section \ref{Asymptotic symmetries in the gauge fixing approach}, we study the gauge fixing approach and provide some examples related to gravity in asymptotically flat and asymptotically (A)dS spacetimes.      

\subsection{Hamiltonian approach}

Some alternative approaches exist that are also powerful in practice. For example, in the Hamiltonian formalism, asymptotically flat \cite{Regge:1974zd} and AdS \cite{Henneaux:1985tv, Brown:1986nw} spacetimes have been studied at spatial infinity. Furthermore, the global BMS group was recently identified at spatial infinity using twisted parity conditions \cite{Henneaux:2018cst, Henneaux:2018gfi , Henneaux:2019yax }. In this framework, the computations are done in a coordinate system making the split between space and time explicit, without performing any gauge fixing. Then, the asymptotic symmetry group is defined as the quotient between the gauge transformations preserving the boundary conditions and the trivial gauge transformations, where trivial means that the associated charges are identically vanishing on the phase space. This definition of the asymptotic symmetry group corresponds to the strong definition in the two first approaches.

\section{Asymptotic symmetries in the gauge fixing approach}
\label{Asymptotic symmetries in the gauge fixing approach}

We now focus on the aforementioned gauge fixing approach of asymptotic symmetries in gauge theories. We illustrate the different definitions and concepts using examples, with a specific focus on asymptotically flat and asymptotically (A)dS spacetimes in four-dimensional general relativity. 

\subsection{Gauge fixing procedure}

\paragraph{Definition \normalfont[Gauge symmetry]} Let us start with a Lagrangian theory in a $n$-dimensional spacetime $M$
\begin{equation}
S[\Phi] = \int_M \mathbf{L}[\Phi, \partial_\mu \Phi, \partial_{\mu}\partial_{\nu} \Phi, \ldots] ,
\label{Lagrangian of the theory}
\end{equation} where $\mathbf{L} = L \mathrm{d}^n x$ is the Lagrangian and $\Phi = (\phi^i)$ are the fields of the theory. A \textit{gauge transformation} is a transformation acting on the fields, and which depends on parameters $F=(f^\alpha)$ that are taken to be arbitrary functions of the spacetime coordinates. We write
\begin{equation}
\begin{split}
\delta_F \Phi &= R[ F] \\
&=R_\alpha f^\alpha + R^\mu_\alpha \partial_\mu f^\alpha + R^{(\mu\nu)}_\alpha \partial_\mu \partial_\nu f^\alpha + \ldots \\
&= \sum_{k\ge 0} R^{(\mu_1\ldots\mu_k)}_\alpha \partial_{\mu_1} \ldots \partial_{\mu_k} f^\alpha \\
\end{split}
\label{transformation of the fields}
\end{equation} the infinitesimal gauge transformation of the fields. In this expression, $R^{(\mu_1\ldots\mu_k)}_\alpha$ are \textit{local functions}, namely functions of the coordinates, the fields, and their derivatives. The gauge transformation is a \textit{symmetry} of the theory if, under \eqref{transformation of the fields}, the Lagrangian transforms as
\begin{equation}
\delta_F \mathbf{L} = \mathrm{d} \mathbf{B}_F,
\label{symmetry}
\end{equation} where $\mathbf{B}_F = B_F^\mu (\mathrm{d}^{n-1}x)_\mu$.  

\paragraph{Examples} We illustrate this definition by providing some examples. First, consider classical vacuum electrodynamics
\begin{equation}
S[A] = \int_M \mathbf{F} \wedge \star \mathbf{F},
\label{electrodynamics}
\end{equation} where $\mathbf{F} = \mathrm{d} \mathbf{A}$ and $\mathbf{A}$ is a $1$-form. It is straightforward to check that the gauge transformation $\delta_\alpha \mathbf{A} = \mathrm{d} \alpha$, where $\alpha$ is an arbitrary function of the coordinates, is a symmetry of the theory.

%Similarly, consider the Yang-Mills theory with Lagrangian \eqref{electrodynamics} and $F= dA + \frac{1}{2}[ A , A ]$, where $A$ is a Lie algebra-valued $1$-form and the bracket is a graded Lie bracket. We can check that the gauge transformation $\delta_\epsilon A = d\epsilon + [\epsilon , A]$, where $\epsilon$ is an arbitrary function of the coordinates, is a symmetry of the theory.

Now, consider the general relativity theory
\begin{equation}
S[g] = \frac{1}{16\pi G} \int_M (R-2 \Lambda) \sqrt{-g} \mathrm{d}^n x ,
\end{equation} where $R$ and $\sqrt{-g}$ are the scalar curvature and the square root of minus the determinant associated with the metric $g_{\mu \nu}$ respectively, and $\Lambda$ is the cosmological constant. It can be checked that the gauge transformation $\delta_\xi g_{\mu\nu} = \mathcal{L}_\xi g_{\mu \nu} = \xi^\rho \partial_\rho g_{\mu\nu} + g_{\mu\rho} \partial_\nu \xi^\rho +  g_{\rho \nu} \partial_\mu \xi^\rho$, where $\xi^\mu$ is a vector field generating a diffeomorphism, is a symmetry of the theory.

Notice that in these examples, the transformation of the fields \eqref{transformation of the fields} is of the form 
\begin{equation}
\delta_F \Phi =  R_\alpha f^\alpha + R^\mu_\alpha \partial_\mu f^\alpha ,
\label{First order}
\end{equation} namely they involve at most first order derivatives of the parameters. 

\paragraph{Definition \normalfont[Gauge fixing]} Starting from a Lagrangian theory \eqref{Lagrangian of the theory} with gauge symmetry \eqref{transformation of the fields}, the \textit{gauge fixing} procedure involves imposing some algebraic or differential constraints on the fields in order to eliminate (part of) the redundancy in the description of the theory. We write 
\begin{equation}
G[\Phi] = 0
% = G \Phi + G^\mu \partial_\mu \Phi +  G^{(\mu\nu)} \partial_\mu \partial_\nu \Phi + \ldots
\label{generic gauge condition}
\end{equation} a generic gauge fixing condition. This gauge has to satisfy two conditions: 
\begin{itemize}
\item It has to be reachable by a gauge transformation, which means that the number of independent conditions in \eqref{generic gauge condition} is inferior or equal to the number of independent parameters $F = (f^\alpha)$ generating the gauge transformation.
\item It has to use all of the available freedom of the arbitrary functions parametrizing the gauge transformations to reach the gauge\footnote{If the available freedom is not used, we talk about \textit{partial gauge fixing}. In this configuration, there are still some arbitrary functions of the coordinates in the parameters of the residual gauge transformations.}, which means that the number of independent conditions in \eqref{generic gauge condition} is superior or equal to the number of independent parameters $F = (f^\alpha)$ generating the gauge transformations.
\end{itemize}
Considering these two requirements together tells us that the number of independent gauge fixing conditions in \eqref{generic gauge condition} has to be equal to the number of independent gauge parameters $F = (f^\alpha)$ involved in the fields transformation \eqref{transformation of the fields}.

\paragraph{Examples} In electrodynamics, several gauge fixings are commonly used. Let us mention the Lorenz gauge $\partial^\mu A_\mu = 0$, the Coulomb gauge $\partial^i A_i=0$, the temporal gauge $A_0 = 0$, and the axial gauge $A_3 = 0$. As previously discussed, the Lorenz gauge can always be reached by performing a gauge transformation. We can check that the same statement holds for all the other gauge fixings. Notice that these gauge fixing conditions involve only one constraint, as there is only one free parameter $\alpha$ in the gauge transformation. 

%Similar gauge fixings hold in Yang-Mills theory. For example, writing $A_\mu = A^a_\mu T^a$ where $T^a$ are the generators of the $SU(N)$ gauge group, the Lorentz gauge becomes $\partial^\mu A_\mu^a = 0$. The number of independent constraints is equal to the number of independent gauge parameters in the transformation, $\epsilon = \epsilon^a T^a$.

%\textbf{Talk about Gribov problem??}

In gravity, many gauge fixings are also used in practice. For example, the \textit{De Donder (or harmonic) gauge} requires that the coordinates $x^\mu$ be harmonic functions, namely, $\Box x^\mu = \frac{1}{\sqrt{-g}} \partial_\nu ( \sqrt{-g} \partial^\nu x^\mu)= 0$. Notice that the number of constraints, $n$, is equal to the number of independent gauge parameters $\xi^\mu$. This gauge condition is suitable for studying gravitational waves in perturbation theory (see, e.g., \cite{Blanchet:2013haa}).

Another important gauge fixing in configurations where $\Lambda \neq 0$ is the \textit{Fefferman-Graham gauge} \cite{Starobinsky:1982mr, AST_1985__S131__95_0, Skenderis:2002wp, 2007arXiv0710.0919F, Papadimitriou:2010as}. We write the coordinates as $x^\mu = (\rho, x^a)$, where $a=1, \ldots,n-1$ and $\rho$ is an expansion parameter ($\rho = 0$ is at the spacetime boundary, and $\rho > 0$ is in the bulk). It is defined by the following conditions:
\begin{equation}
g_{\rho\rho} = - \frac{(n-1)(n-2)}{2\Lambda \rho^2}, \quad g_{\rho a} = 0
\label{FG gauge}
\end{equation} ($n$ conditions). The coordinate $\rho$ is spacelike for $\Lambda<0$ and timelike for $\Lambda>0$. The most general metric takes the form
\begin{equation}
\mathrm{d}s^2 =- \frac{(n-1)(n-2)}{2\Lambda}\frac{\mathrm{d}\rho^2}{\rho^2} + g_{ab}(\rho,x^c) \mathrm{d}x^a \mathrm{d}x^b. 
\label{fG gauge}
\end{equation}

Finally, the \textit{Bondi gauge} will be relevant for us in the following \cite{Bondi:1962px , Sachs:1962wk , Sachs:1962zza}. This gauge fixing is valid for both $\Lambda = 0$ and $\Lambda \neq 0$ configurations. Writing the coordinates as $(u,r,x^A)$, where $x^A = (\theta_1, \ldots , \theta_{n-2})$ are the transverse angular coordinates on the $(n-2)$-celestial sphere, the Bondi gauge is defined by the following conditions:
\begin{equation}
g_{rr} = 0, \quad g_{rA} = 0, \quad \partial_r \left(\frac{\det g_{AB}}{r^{2(n-2)}} \right) = 0
\label{Bondi gauge}
\end{equation}  ($n$ conditions). These conditions tell us that, geometrically, $u$ labels null hypersurfaces in the spacetime, $x^A$ labels null geodesics inside a null hypersurface, and $r$ is the luminosity distance along the null geodesics. The most general metric takes the form
\begin{equation}
\mathrm{d}s^2 = e^{2\beta} \frac{V}{r} \mathrm{d}u^2 - 2 e^{2\beta}\mathrm{d}u \mathrm{d}r + g_{AB} (\mathrm{d}x^A - U^A \mathrm{d}u)(dx^B - U^B \mathrm{d}u)
\label{bondi gauge}
\end{equation} where $\beta$, $U^A$ and $\frac{V}{r}$ are arbitrary functions of the coordinates, and the $(n-2)$-dimensional metric $g_{AB}$ satisfies the determinant condition in the third equation of \eqref{Bondi gauge}. Let us mention that the Bondi gauge is closely related to the \textit{Newman-Unti gauge} \cite{Newman:1962cia, Barnich:2011ty} involving only algebraic conditions:
\begin{equation}
g_{rr} = 0, \quad g_{rA} = 0, \quad g_{ru} = -1
\end{equation} ($n$ conditions).

\paragraph{Definition \normalfont[Residual gauge transformation]} After having imposed a gauge fixing as in equation \eqref{generic gauge condition}, there usually remain some \textit{residual gauge transformations}, namely gauge transformations preserving the gauge fixing condition. Formally, the residual gauge transformations with generators $F$ have to satisfy $\delta_F G[\Phi]  = 0$. They are local functions parametrized as $F = F(a)$, where the parameters $a$ are arbitrary functions of $(n-1)$ coordinates.
%\begin{equation}
%\begin{split}
%\delta_F G[\Phi]  
%&=  G \delta_F \Phi + G^\mu \partial_\mu \delta_F \Phi +  G^{(\mu\nu)} \partial_\mu %\partial_\nu \delta_F \Phi + \ldots\\
%&= G[\delta_F \Phi ]\\
%&= 0
%\end{split}
%\end{equation}

\paragraph{Examples} Consider the Lorenz gauge $\partial^\mu A_\mu=0$ in electrodynamics. As we discussed earlier, the residual gauge transformations for the Lorenz gauge are the gauge transformations $\delta_\alpha A_\mu = \partial_\mu \alpha$, where $\partial^\mu\partial_\mu \alpha = 0$. 

Similarly, consider the Fefferman-Graham gauge \eqref{FG gauge} in general relativity with $\Lambda \neq 0$. The residual gauge transformations generated by $\xi^\mu$ have to satisfy $\mathcal{L}_\xi g_{\rho \rho} = 0$ and $\mathcal{L}_\xi g_{\rho a} = 0$. The solutions to these equations are given by
\begin{equation}
\xi^\rho = \sigma (x^a) \rho , \quad \xi^a = \xi^a_0 (x^b) + \frac{(n-1)(n-2)}{2\Lambda} \partial_b \sigma \int^\rho_0 \frac{\mathrm{d} \rho'}{\rho'} g^{ab}(\rho',x^c) .
\label{residual FG}
\end{equation} These solutions are parametrized by $n$ arbitrary functions $\sigma$ and $\xi^a_0$ of $(n-1)$ coordinates $x^a$. 

In the Bondi gauge \eqref{Bondi gauge}, the residual gauge transformations generated by $\xi^\mu$ have to satisfy $\mathcal{L}_\xi g_{rr} = 0$, $\mathcal{L}_\xi g_{rA} = 0$ and $g^{AB} \mathcal{L}_\xi g_{AB} = 4 \omega$, where $\omega$ is an arbitrary function of $(u,x^A)$. The solutions to these equations are given by
\begin{equation}
\begin{split}
\xi^u &= f, \\
\xi^A &= Y^A + I^A, \quad I^A = -\partial_B f \int_r^\infty \mathrm{d}r'  (e^{2 \beta} g^{AB}),\\
\xi^r &= - \frac{r}{n-2} (\mathcal{D}_A Y^A - 2 \omega + \mathcal{D}_A I^A - \partial_B f U^B + \frac{1}{2} f g^{-1} \partial_u g) ,\\
\end{split}
\label{eq:xir}
\end{equation} 
where $\partial_r f = 0 = \partial_r Y^A$, and $g= \det (g_{AB})$ \cite{Barnich:2013sxa}. The covariant derivative $\mathcal{D}_A$ is associated with the $(n-2)$-dimensional metric $g_{AB}$. The residual gauge transformations are parametrized by the $n$ functions $\omega$, $f$ and $Y^A$ of $(n-1)$ coordinates $(u,x^A)$.

\subsection{Boundary conditions}

\paragraph{Definition \normalfont[Boundary conditions]} Once a gauge condition \eqref{generic gauge condition} has been fixed, we can impose \textit{boundary conditions} for the theory by requiering some constraints on the fields in a neighbourhood of a given spacetime region. Most of those boundary conditions are fall-off conditions on the fields in the considered asymptotic region\footnote{Notice that the asymptotic region could be taken not only at (spacelike, null or timelike) infinity, but also in other spacetime regions, such as near a black hole horizon \cite{Hawking:2016msc, Hawking:2016sgy , Haco:2018ske , Haco:2019ggi , Donnay:2015abr , Donnay:2016ejv , Donnay:2019jiz , Donnay:2019zif , Grumiller:2019fmp}.}, or conditions on the leading functions in the expansion. This choice of boundary conditions is motivated by the physical model that we want to consider. A set of boundary conditions is usually considered to be interesting if it provides non-trivial asymptotic symmetry group and solution space, exhibiting interesting properties for the associated charges (finite, generically non-vanishing, integrable and conserved; see below). If the boundary conditions are too strong, the asymptotic symmetry group will be trivial, with vanishing surface charges. Furthermore, the solution space will not contain any solution of interest. If they are too weak, the associated surface charges will be divergent. Consistent and interesting boundary conditions should therefore be located between these two extreme situations.

\paragraph{Examples} Let us give some examples of boundary conditions in general relativity theory. Many examples of boundary conditions for other gauge theories can be found in the literature (see e.g. \cite{Strominger:2013lka , He:2014cra , Afshar:2018apx , Detournay:2018cbf , Barnich:2013sxa , Capone:2019aiy , Lu:2019jus}). 

Let us consider the \textit{Bondi gauge} defined in equation \eqref{Bondi gauge} in dimension $n \ge 3$. There exist several definitions of \textit{asymptotic flatness at null infinity} ($r\to \infty$) in the literature. For all of them, we require the following preliminary boundary conditions on the functions of the metric \eqref{bondi gauge} in the asymptotic region $r\to\infty$:
\begin{equation}
\beta = o(1), \quad \frac{V}{r} = o(r^{2}), \quad U^A = o(1), \quad g_{AB} = r^2 q_{AB} + r C_{AB} + D_{AB} + \mathcal{O}(r^{-1}) ,
\label{asymp flat 1}
\end{equation} where $q_{AB}$, $C_{AB}$ and $D_{AB}$ are $(n-2)$-dimensional symmetric tensors, which are functions of $(u,x^A)$. Notice in particular that $q_{AB}$ is kept free at this stage. 

A first definition of asymptotic flatness at null infinity (AF1) is a sub-case of \eqref{asymp flat 1}. In addition to all these fall-off conditions, we require the transverse boundary metric $q_{AB}$ to have a fixed determinant, namely,
\begin{equation}
\sqrt{q} = \sqrt{\bar{q}} ,
\label{asymp flat 1 prime}
\end{equation} where $\bar{q}$ is a fixed volume element (which may possibly depend on time) on the $(n-2)$-dimensional transverse space \cite{Campiglia:2014yka, Campiglia:2015yka, Compere:2018ylh , Flanagan:2019vbl}. 

A second definition of asymptotic flatness at null infinity (AF2) is another sub-case of the definition \eqref{asymp flat 1}. All the conditions are the same, except that we require that the transverse boundary metric $q_{AB}$ be conformally related to the unit $(n-2)$-sphere metric, namely,
\begin{equation}
q_{AB} = e^{2\varphi} \mathring{q}_{AB} ,
\label{asymp flat 2}
\end{equation} where $\mathring{q}_{AB}$ is the unit $(n-2)$-sphere metric \cite{Barnich:2010eb}. Note that for $n = 4$, this condition can always be reached by a coordinate transformation, since every metric on a two dimensional surface is conformally flat (but even in this case, as we will see below, this restricts the form of the symmetries). 

A third definition of asymptotic flatness at null infinity (AF3), which is the historical one \cite{Bondi:1962px , Sachs:1962wk , Sachs:1962zza}, is a sub-case of the second definition \eqref{asymp flat 2}. We require \eqref{asymp flat 1} and we demand that the transverse boundary metric $q_{AB}$ be the unit $(n-2)$-sphere metric, namely,
\begin{equation}
q_{AB} = \mathring{q}_{AB} .
\label{asymp flat 3}
\end{equation} Note that this definition of asymptotic flatness is the only one that has the property to be \textit{asymptotically Minkowskian}, that is, for $r\to \infty$, the leading orders of the spacetime metric \eqref{bondi gauge} tend to the Minkowski line element $\mathrm{d}s^2 = -\mathrm{d}u^2 -2 \mathrm{d}u \mathrm{d}r + r^2 \mathring{q}_{AB} \mathrm{d}x^A \mathrm{d}x^B$. 

Let us now present several definitions of \textit{asymptotically (A)dS spacetimes} in both the \textit{Fefferman Graham gauge} \eqref{FG gauge} and \textit{Bondi gauge} \eqref{Bondi gauge}. A preliminary boundary condition, usually called the asymptotically locally (A)dS condition, requires the following conditions on the functions of the Fefferman-Graham metric \eqref{fG gauge}:
\begin{equation}
g_{ab} = \mathcal{O}(\rho^{-2})
\label{as ads}
\end{equation} or, equivalently, $g_{ab} = \rho^{-2} g_{ab}^{(0)} + o(\rho^{-2})$. Notice that the $(n-1)$-dimensional boundary metric $g^{(0)}_{ab}$ is kept free in this preliminary set of boundary conditions, thus justifying the adjective ``locally" \cite{Fischetti:2012rd}. In the Bondi gauge, as we will see below, these fall-off conditions are equivalent to demand that
\begin{equation}
g_{AB} = \mathcal{O}(r^2)
\label{as ads Bondi}
\end{equation} or, equivalently, $g_{AB} = r^2 q_{AB} + o(r^2)$. 

A first definition of asymptotically (A)dS spacetime (AAdS1) is a sub-case of the definition \eqref{as ads}. In addition to these fall-off conditions, we demand the following constraints on the $(n-1)$-dimensional boundary metric $g^{(0)}_{ab}$ :
\begin{equation}
g_{tt}^{(0)} = \frac{2\Lambda}{(n-1)(n-2)}, \quad g_{tA}^{(0)} = 0, \quad \det(g_{ab}^{(0)}) = \frac{2 \Lambda}{(n-1)(n-2)} \bar{q} ,
\label{boundary gauge fixing FG}
\end{equation} where $\bar{q}$ is a fixed volume form for the transverse $(n-2)$-dimensional space (which may possibly depend on $t$) \cite{Compere:2019bua}. In the Bondi gauge, the boundary conditions \eqref{boundary gauge fixing FG} translate into  
\begin{equation}
\beta = o(1), \quad \frac{V}{r} = \frac{2r^2 \Lambda}{(n-1)(n-2)} + o(r^2), \quad U^A = o(1), \quad  \sqrt{q} = \sqrt{\bar{q}} .
\label{boundary gauge fixing Bondi}
\end{equation} Notice the similarity of these conditions to the definition (AF1) (equations \eqref{asymp flat 1} and \eqref{asymp flat 1 prime}) of asymptotically flat spacetime.

A second definition of asymptotically AdS spacetime\footnote{This choice is less relevant for asymptotically dS spacetimes, since it strongly restricts the Cauchy problem and the bulk spacetime dynamics \cite{Ashtekar:2014zfa , Ashtekar:2015lla}.} (AAdS2) is a sub-case of the definition \eqref{as ads}. We require the same conditions as in the preliminary boundary condition \eqref{as ads}, except that we demand that the $(n-1)$-dimensional boundary metric $g^{(0)}_{ab}$ be fixed \cite{Henneaux:1985tv}. These conditions are called \textit{Dirichlet boundary conditions}. One usually chooses the cylinder metric as the boundary metric, namely,
\begin{equation}
g_{ab}^{(0)} \mathrm{d}x^a \mathrm{d}x^b =  \frac{2 \Lambda }{(n-1)(n-2)} \mathrm{d}t^2 + \mathring{q}_{AB} \mathrm{d}x^A \mathrm{d}x^B ,
\label{BC Dirichlet}
\end{equation} where $\mathring{q}_{AB}$ are the components of the unit $(n-2)$-sphere metric (as in the Bondi gauge, the upper case indices $A,B, \ldots$ run from $3$ to $n$, and $x^a = (t,x^A)$). In the Bondi gauge, the boundary conditions \eqref{BC Dirichlet} translate into 
\begin{equation}
\beta = o(1), \quad \frac{V}{r} = \frac{2r^2 \Lambda}{(n-1)(n-2)} + o(r^2), \quad U^A = o (1), \quad q_{AB} = \mathring{q}_{AB} .
\label{BC Dirichlet Bondi}
\end{equation} Notice the similarity of these conditions to the definition (AF3) (equations \eqref{asymp flat 1} and \eqref{asymp flat 3}) of asymptotically flat spacetime. 

\begin{figure}[h!]
	\centering
	\vspace{10pt}
	\begin{tabular}{ccc}
	\includegraphics[width=0.3\textwidth]{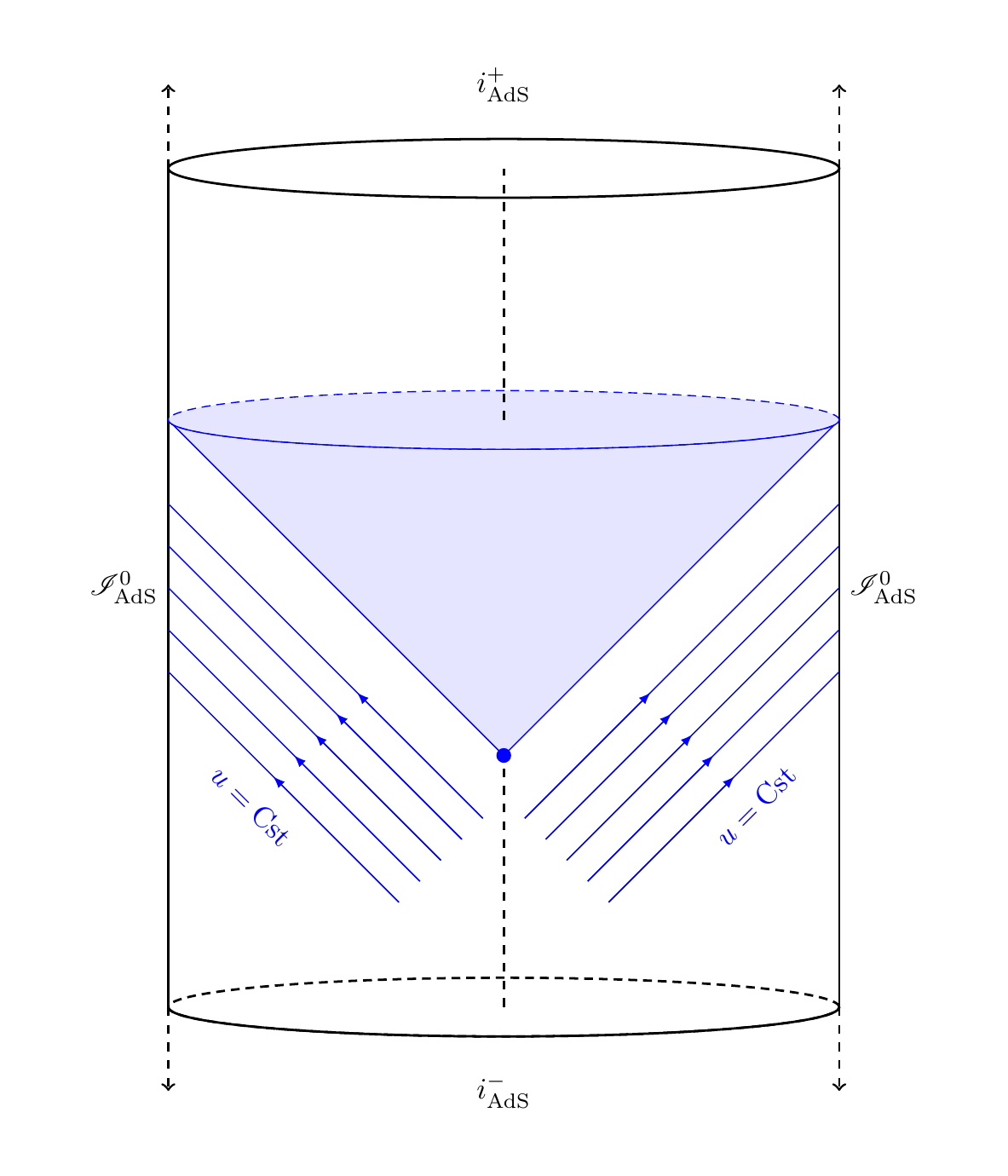}\label{Fig:AdS} & \includegraphics[width=0.3\textwidth]{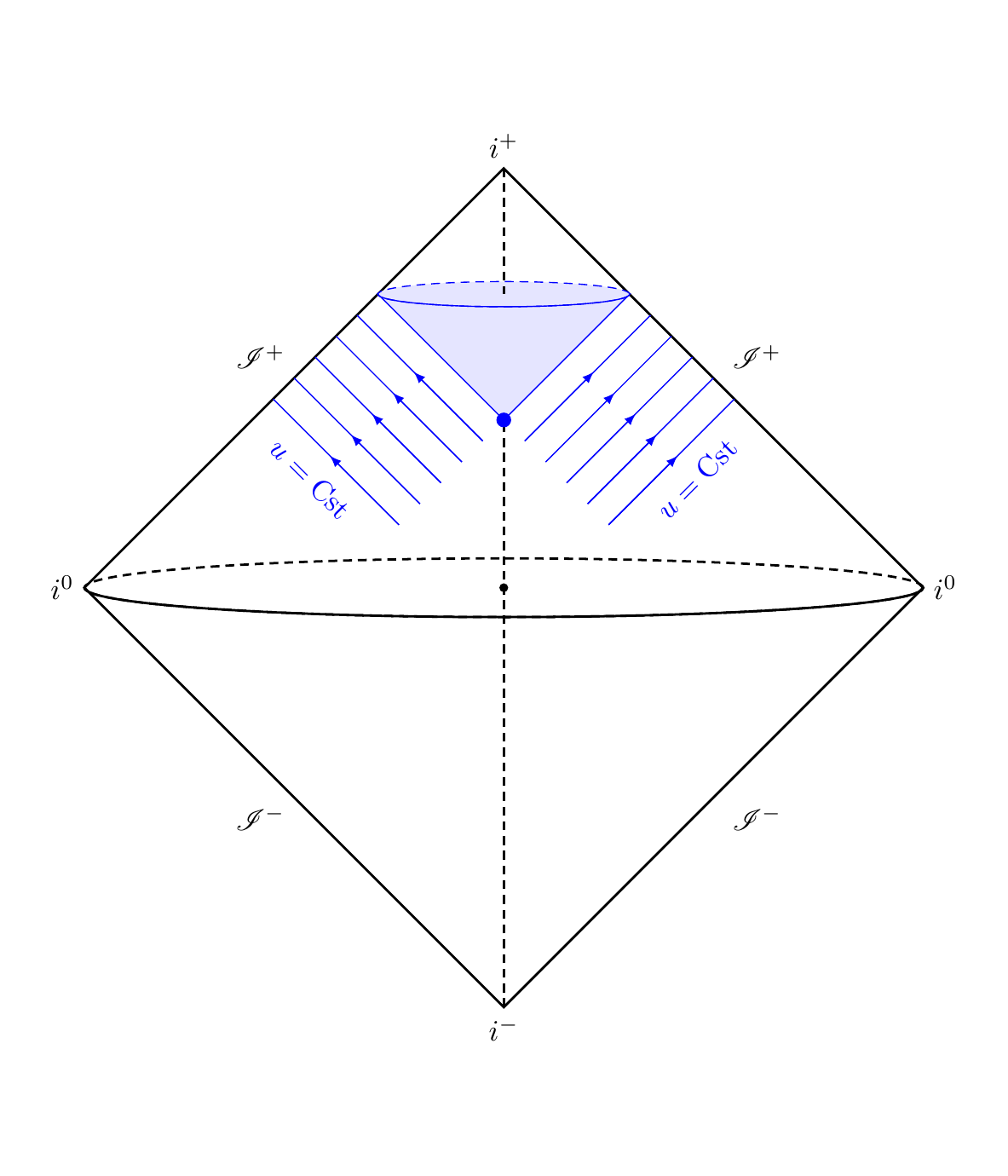}\label{Fig:Flat} & \includegraphics[width=0.3\textwidth]{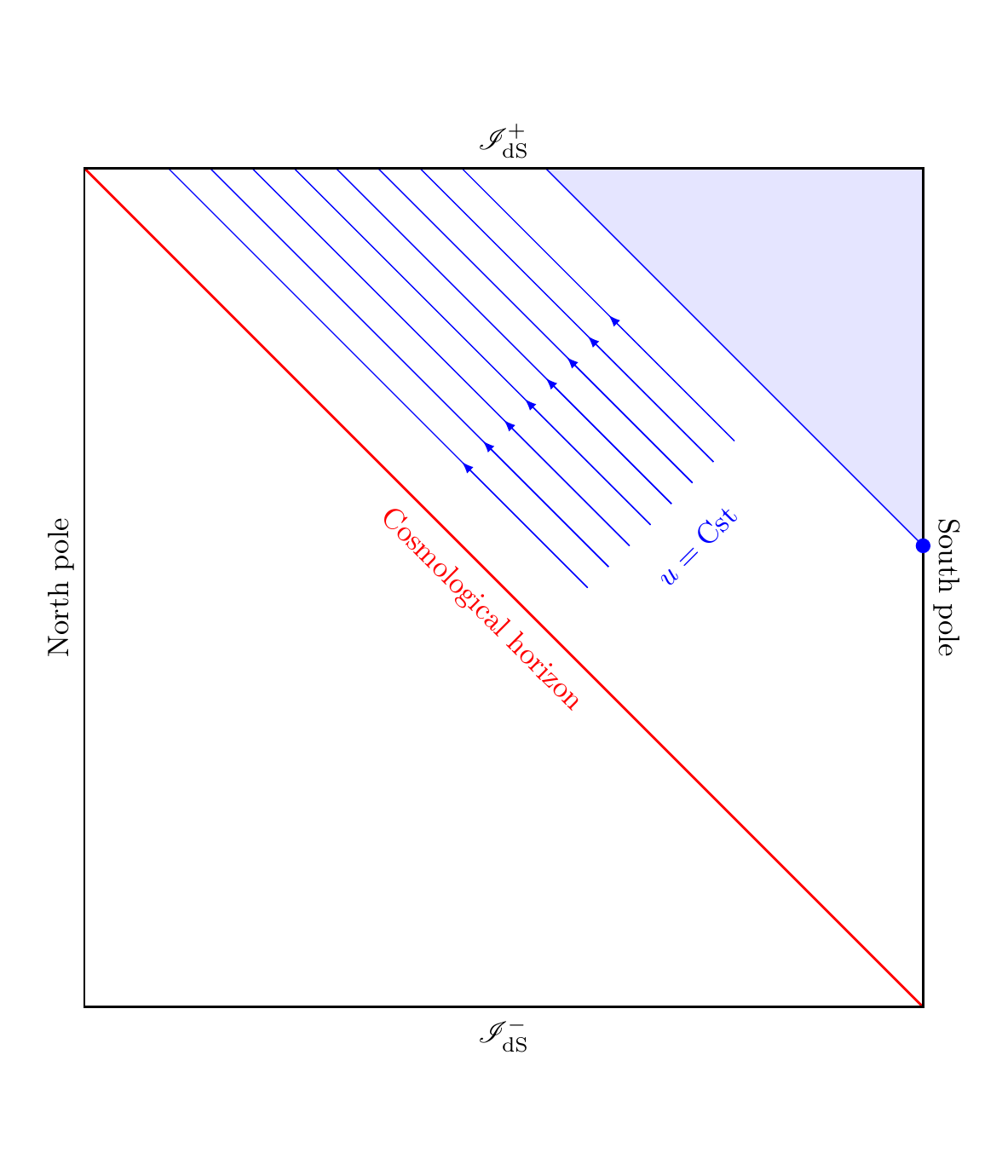}\label{Fig:dS} \\  
	AdS case $(\Lambda<0)$. & Flat case $(\Lambda=0)$. & dS case $(\Lambda>0)$.  
	\end{tabular} 
	\vspace{20pt}
	\caption{Bondi gauge for any $\Lambda$.}
	\label{Fig:FIGUREE}
\end{figure}

As we see it, the Bondi gauge is well-adapted for each type of asymptotics (see figure \ref{Fig:FIGUREE}), while the Fefferman-Graham gauge is only defined in asymptotically (A)dS spacetimes. 

%Let us notice that other sets of boundary conditions, specifically relevant for asymptotically AdS spactimes, will be introduced later in the text. 

%\begin{center}
%
%\begin{figure}[!h]
%   \begin{minipage}[t]{0.30\linewidth}
%     \centering
%     \includegraphics[width=\linewidth]{AdS.pdf}
%     \caption*{Asymptotically AdS case ($\Lambda <0$).}\label{Fig:AdS}
%   \end{minipage}
%   \begin{minipage}[t]{0.30\textwidth}
%     \centering
%     \includegraphics[width=\linewidth]{Flat.pdf}
%     \caption*{Asymptotically flat case ($\Lambda = 0$).}\label{Fig:Flat}
%   \end{minipage}
%   \begin{minipage}[t]{0.30\textwidth}
%     \centering
%     \includegraphics[width=\linewidth]{dS.pdf}
%     \caption*{Asymptotically dS case ($\Lambda > 0$).}\label{Fig:dS}
%   \end{minipage}
%   \label{FIGUREE}
%\end{figure}
%
%
%\end{center}

%A third definition of asymptotically AdS spacetime is a sub-case of the definition \eqref{as ads}. We require the same fall-off condition, but we demand that the order $(n-1)$ in the expansion
%\begin{equation}
%g_{ab} = \rho^{-2} g_{ab}^{(0)} + \rho^{-1} g_{ab}^{(1)} + \ldots + \rho^{n-3} g_{ab}^{(n-1)} + \ldots 
%\label{BC Newmann}
%\end{equation} namely $g_{ab}^{(n-1)}$, is fixed (e.g. set to zero), while $g_{ab}^{(0)}$ is kept arbitrary. These are called Newmann boundary conditions \cite{Compere:2008us}. In the Bondi gauge, the boundary conditions \eqref{BC Newmann} translate into 
%\begin{equation}
%\beta = o(1), \quad \frac{V}{r} = \frac{2r^2 \Lambda}{(n-1)(n-2)} + o(r^2), \quad U^A = o (1), \quad q_{AB} = \mathring{q}_{AB}
%\end{equation}

\subsection{Solution space}
\label{Solution space}

\paragraph{Definition \normalfont[Solution space]} Given a gauge fixing \eqref{generic gauge condition} and boundary conditions, a \textit{solution} of the theory is a field configuration $\tilde{\Phi}$ satisfying $G[\tilde{\Phi}]=0$, the boundary conditions, and the \textit{Euler Lagrange-equations}
\begin{equation}
\left. \frac{\delta \mathbf{L}}{\delta \Phi}\right\vert_{\tilde{\Phi}} = 0 ,
\end{equation} where the \textit{Euler-Lagrange derivative} is defined in equation \eqref{Euler lagrange def}. The set of all solutions of the theory is called the \textit{solution space}. It is parametrized as $\tilde{\Phi} = \tilde{\Phi}(b)$, where the parameters $b$ are arbitrary functions of $(n-1)$ coordinates.

\paragraph{Examples} We now provide some examples of solution spaces of four-dimensional general relativity in different gauge fixings. We first consider the \textit{Fefferman-Graham gauge} in asymptotically (A)dS$_4$ spacetimes with preliminary boundary conditions \eqref{as ads}. Solving the Einstein equations
\begin{equation}
G_{\mu\nu} + \Lambda g_{\mu\nu} = 0 ,
\end{equation} we obtain the following analytic fall-offs:
\begin{equation}
g_{ab} = \rho^{-2} g_{ab}^{(0)} + \rho^{-1} g_{ab}^{(1)} + g_{ab}^{(2)} + \rho  g_{ab}^{(3)} + \mathcal{O}(\rho^{2}) ,
\end{equation} where $g^{(i)}_{ab}$ are functions of $x^a$ \cite{Starobinsky:1982mr, AST_1985__S131__95_0, Skenderis:2002wp, 2007arXiv0710.0919F, Papadimitriou:2010as}. The only free data in this expansion are $g_{ab}^{(0)}$ and $g_{ab}^{(3)}$. All the other coefficients are determined in terms of these free data. Following the holographic dictionary, we call $g_{ab}^{(0)}$ the boundary metric and we define 
\begin{equation}
T_{ab} = \frac{\sqrt{3|\Lambda|}}{16\pi G} g_{ab}^{(3)}
\end{equation} as the stress energy tensor. From the Einstein equations, we have
\begin{equation}
g_{ab}^{(0)} T^{ab} = 0, \quad D_a^{(0)} T^{ab} = 0 ,
\label{stress energy tensor cond}
\end{equation} where $D_a^{(0)}$ is the covariant derivative with respect to $g_{ab}^{(0)}$. In summary, the solution space of general relativity in the Fefferman-Graham gauge with the preliminary boundary condition \eqref{as ads} is parametrized by the set of functions 
\begin{equation}
\{g_{ab}^{(0)}, T_{ab} \}_{\Lambda \neq 0} ,
\label{most general sol space fg}
\end{equation} where $T_{ab}$ satisfies \eqref{stress energy tensor cond} ($11$ functions).

Now, for the restricted set of boundary conditions \eqref{boundary gauge fixing FG}, that is, (AAdS1), the solution space reduces to 
\begin{equation}
\{g_{AB}^{(0)}, T_{ab} \}_{\Lambda \neq 0} ,
\end{equation} where $g_{AB}^{(0)}$ has a fixed determinant and $T_{ab}$ satisfies \eqref{stress energy tensor cond} ($7$ functions). Finally, for Dirichlet boundary conditions \eqref{BC Dirichlet} (AAdS2), the solution space reduces to 
\begin{equation}
\{T_{ab} \}_{\Lambda \neq 0} ,
\label{solution space dirichlet}
\end{equation} where $T_{ab}$ satisfies \eqref{stress energy tensor cond} (5 functions).

Let us now consider the \textit{Bondi gauge} in asymptotically (A)dS$_4$ spacetimes with preliminary boundary condition \eqref{as ads Bondi}. From the Fefferman-Graham theorem and the gauge matching between Bondi and Fefferman-Graham that is described in appendix \ref{Diffeomorphism between Bondi and Fefferman-Graham gauges} (see also \cite{Poole:2018koa , Compere:2019bua}), we know that the functions appearing in the metric admit an analytic expansion in powers of $r$. In particular, we can write 
\begin{equation}
g_{AB} = r^2 q_{AB} + r C_{AB} + D_{AB} + \frac{1}{r} E_{AB} + \frac{1}{r^2} F_{AB} +\mathcal{O}(r^{-3}) ,
\label{preliminary boundary condition}
\end{equation} where $q_{AB}$, $C_{AB}$, $D_{AB}$, $E_{AB}$, $F_{AB}$, $\ldots$ are functions of $(u, x^A)$. The determinant condition defining the Bondi gauge and appearing in the third equation of \eqref{Bondi gauge} implies $g^{AB}\partial_r g_{AB}=4/r$, which imposes successively that $\det (g_{AB}) = r^4 \det (q_{AB})$, $q^{AB} C_{AB} = 0$ and
\begin{equation}
\begin{split}
&D_{AB} = \frac{1}{4} q_{AB} C^{CD} C_{CD} + \mathcal{D}_{AB} (u,x^C),  \\
&E_{AB} = \frac{1}{2} q_{AB} \mathcal{D}_{CD}C^{CD} + \mathcal{E}_{AB} (u,x^C), \\
&F_{AB} = \frac{1}{2} q_{AB} \Big[ C^{CD}\mathcal{E}_{CD} + \frac{1}{2} \mathcal{D}^{CD}\mathcal{D}_{CD} - \frac{1}{32} (C^{CD}C_{CD})^2 \Big] + \mathcal{F}_{AB}(u,x^C),
\end{split}
\end{equation}
with $q^{AB} \mathcal{D}_{AB} = q^{AB} \mathcal{E}_{AB} = q^{AB} \mathcal{F}_{AB} = 0$ (indices are lowered and raised with the metric $q_{AB}$ and its inverse). We now sketch the results obtained by solving the Einstein equations
\begin{equation}
G_{\mu\nu} + \Lambda g_{\mu\nu} = 0
\end{equation} for $\Lambda \neq 0$ (we follow \cite{Compere:2019bua, Poole:2018koa}; see also \cite{Mao:2019ahc} for the Newman-Penrose version). The component $(rr)$ gives the following radial constraints on the Bondi functions:
\begin{align}
\beta(u,r,x^A) &= \beta_0 (u,x^A) + \frac{1}{r^2} \Big[ -\frac{1}{32} C^{AB} C_{AB} \Big] + \frac{1}{r^3} \Big[ -\frac{1}{12} C^{AB} \mathcal{D}_{AB} \Big] \label{eq:EOM_beta} \\
&\qquad + \frac{1}{r^4}\Big[ - \frac{3}{32} C^{AB}\mathcal{E}_{AB} - \frac{1}{16} \mathcal{D}^{AB}\mathcal{D}_{AB} + \frac{1}{128} (C^{AB}C_{AB})^2 \Big] + \mathcal{O}(r^{-5}). \nonumber
\end{align} where $\beta_0 (u,x^A)$ is an arbitrary function. The component $(rA)$ yields 
\begin{equation}
\begin{split}
U^A = \,\, & U^A_0(u,x^B) +\overset{(1)}{U^A}(u,x^B) \frac{1}{r} + \overset{(2)}{U^A}(u,x^B) \frac{1}{r^2} \\
&+ \overset{(3)}{U^A}(u,x^B) \frac{1}{r^3} + \overset{(\text{L}3)}{U^A}(u,x^B) \frac{\ln r}{r^3} + o(r^{-3})
\end{split} \label{eq:EOM_UA}
\end{equation}
with
\begin{eqnarray}
\overset{(1)}{U^A}(u,x^B)\hspace{-6pt} &=&\hspace{-6pt} 2 e^{2\beta_0} \partial^A \beta_0 ,\nonumber \\
\overset{(2)}{U^A}(u,x^B)\hspace{-6pt} &=&\hspace{-6pt} - e^{2\beta_0} \Big[ C^{AB} \partial_B \beta_0 + \frac{1}{2} D_B C^{AB} \Big], \nonumber\\
\overset{(3)}{U^A}(u,x^B)\hspace{-6pt} &=& \hspace{-6pt}- \frac{2}{3} e^{2\beta_0} \Big[ N^A - \frac{1}{2} C^{AB} D^C C_{BC} +   (\partial_B \beta_0 - \frac{1}{3} D_B) \mathcal{D}^{AB} - \frac{3}{16} C_{CD}C^{CD} \partial^A \beta_0  \Big], \nonumber\\
\overset{(\text{L}3)}{U^A}(u,x^B) \hspace{-6pt}&=&\hspace{-6pt} -\frac{2}{3}e^{2\beta_0}D_B \mathcal{D}^{AB}. \label{eq:EOM_UA2}
\end{eqnarray}
In these expressions, $U^A_0(u,x^B)$ and $N^A(u,x^B)$ are arbitrary functions. We call $N^A$ the \textit{angular momentum aspect}. Notice that, at this stage, logarithmic terms are appearing in the expansion \eqref{eq:EOM_UA}. However, we will see below that these terms vanish for $\Lambda \neq 0$. The component $(ru)$ leads to
\begin{align}
\frac{V}{r} = &\frac{\Lambda}{3} e^{2\beta_0} r^2 - r (l + D_A U^A_0) \label{eq:EOMVr} \\
&- e^{2\beta_0} \Big[ \frac{1}{2}\Big( R[q] + \frac{\Lambda}{8}C_{AB} C^{AB} \Big) + 2 D_A \partial^A \beta_0 + 4 \partial_A \beta_0 \partial^A \beta_0 \Big] - \frac{2  M}{r} + o(r^{-1}) , \nonumber 
\end{align}
where $l = \partial_u \ln \sqrt{q}$, $R[q]$ is the scalar curvature associated with the metric $q_{AB}$ and $ M(u,x^A)$ is an arbitrary function called the \textit{Bondi mass aspect}. Afterwards, we solve the components $(AB)$ of the Einstein equations order by order, thereby providing us with the constraints imposed on each order of $g_{AB}$. The leading order $\mathcal{O}(r^{-1})$ of that equation yields to
\begin{equation}
\frac{\Lambda}{3} C_{AB} = e^{-2\beta_0} \Big[ (\partial_u - l) q_{AB} + 2 D_{(A} U^0_{B)} - D^C U^0_C q_{AB} \Big].
\label{eq:CAB}
\end{equation} Going to $\mathcal{O}(r^{-2})$, we get
\begin{equation}
\frac{\Lambda}{3} \mathcal{D}_{AB} = 0,\label{eq:DAB}
\end{equation}
which removes the logarithmic term in \eqref{eq:EOM_UA} for $\Lambda \neq 0$ (but not for $\Lambda = 0$). The condition at the next order $\mathcal{O}(r^{-3})$
\begin{equation}
\partial_u \mathcal{D}_{AB} + U_0^C D_C \mathcal{D}_{AB} + 2 \mathcal{D}_{C(A} D_{B)}U_0^C = 0
\end{equation}
is trivial for $\Lambda \neq 0$. Using an iterative argument as in \cite{Poole:2018koa}, we now make the following observation. If we decompose $g_{AB} = r^2 \sum_{n\geq 0} g_{AB}^{(n)} r^{-n}$, we see that the iterative solution of the components $(AB)$ of the Einstein equations organizes itself as $\Lambda g_{AB}^{(n)} = \partial_u g_{AB}^{(n-1)} + (...)$ at order $\mathcal{O}(r^{-n})$, $n\in\mathbb{N}_0$. Accordingly, the form of $\mathcal{E}_{AB}$ should have been fixed by the equation found at $\mathcal{O}(r^{-3})$; however, this is not the case, since both contributions of $\mathcal{E}_{AB}$ cancel between $G_{AB}$ and $\Lambda g_{AB}$. Moreover, the equation $\Lambda g_{AB}^{(4)} = \partial_u g_{AB}^{(3)} + (...)$ at the next order turns out to be a constraint for $g_{AB}^{(4)} \sim \mathcal{F}_{AB}$, determined with other subleading data such as $C_{AB}$ or $\partial_u g_{AB}^{(3)} \sim \partial_u \mathcal{E}_{AB}$. It shows that $\mathcal{E}_{AB}$ is a set of two free data on the boundary, built up from two arbitrary functions of $(u,x^A)$. Morover, it indicates that no more data exist to be uncovered for $\Lambda \neq 0$. Finally, the components $(uu)$ and $(uA)$ of the Einstein equations give some evolution constraints with respect to the $u$ coordinate for the Bondi mass aspect $M$ and the angular momentum aspect $N^A$. We will not describe these equations explicitly here (see \cite{Poole:2018koa , Compere:2019bua}).

In summary, the solution space for general relativity in the Bondi gauge with the preliminary boundary condition \eqref{preliminary boundary condition} and $\Lambda \neq 0$ is parametrized by the set of functions
\begin{equation}
\{ \beta_0, U^A_0, q_{AB}, \mathcal{E}_{AB}, M, N^A \}_{\Lambda \neq 0}
\label{most general sol space bondi}
\end{equation} ($11$ functions), where $M$ and $N^A$ have constrained evolutions with respect to the $u$ coordinate. Therefore, the characteristic initial value problem is well-defined when the following data are given: $\beta_0(u, x^C)$, $U^A_0(u, x^C)$, $\mathcal{E}_{AB}(u, x^C)$, $q_{AB}(u, x^C)$, $M(u_0, x^C)$ and $N^A(u_0, x^C)$, where $u_0$ is a fixed value of $u$.

Notice that for the boundary conditions \eqref{boundary gauge fixing Bondi} (AAdS1), the solution space reduces to 
\begin{equation}
\{q_{AB}, \mathcal{E}_{AB}, M, N^A \}_{\Lambda \neq 0} ,
\end{equation} where $M$ and $N^A$ have constrained evolutions with respect to the $u$ coordinate, and $q_{AB}$ has a fixed determinant \cite{Compere:2019bua} ($7$ functions). Finally, for the Dirichlet boundary conditions \eqref{BC Dirichlet Bondi} (AAdS2), the solution space finally reduces to 
\begin{equation}
\{ \mathcal{E}_{AB}, M, N^A \}_{\Lambda \neq 0} ,
\end{equation} where $M$ and $N^A$ have constrained evolutions with respect to the $u$ coordinate (5 functions). 

Let us finally discuss the \textit{Bondi gauge} in asymptotically flat spacetimes \cite{Bondi:1962px , Sachs:1962wk , Sachs:1962zza , Barnich:2010eb, Flanagan:2015pxa , Compere:2018ylh , Compere:2019bua}. We first consider the preliminary boundary conditions \eqref{asymp flat 1}. From the previous analysis of solution space with $\Lambda \neq 0$, we can readily obtain the solution space with $\Lambda =0$, that is, the solution of
\begin{equation}
G_{\mu\nu} = 0 ,
\end{equation} by taking the flat limit $\Lambda \to 0$. The radial constraints \eqref{eq:EOM_beta}, \eqref{eq:EOM_UA2} and \eqref{eq:EOMVr} are still valid by setting to zero $\beta_0$, $U^A_0$ (see equation \eqref{asymp flat 1}) and all the terms proportional to $\Lambda$. Furthermore, by the same procedure, the constraint equation \eqref{eq:CAB} becomes  
\begin{equation}
(\partial_u - l) q_{AB} = 0 .
\label{time evolution constraint}
\end{equation} Therefore, the asymptotic shear $C_{AB}$ becomes unconstrained, and the metric $q_{AB}$ gets a time evolution constraint. Similarly, the equation \eqref{eq:DAB} becomes trivial and $\mathcal{D}_{AB}$ is not constrained at this order. In particular, this allows for the existence of logarithmic terms in the Bondi expansion (see equation \eqref{eq:EOM_UA}). One has to impose the additional condition $D^A \mathcal{D}_{AB} = 0$ to make these logarithmic terms disappear. Finally, one can see that for $\Lambda = 0$, the subleading orders of the components $(AB)$ of the Einstein equations impose time evolution constraints on $\mathcal{D}_{AB}$, $\mathcal{E}_{AB}$, $\ldots$ , but this infinite tower of functions is otherwise unconstrained and they become free parameters of the solution space. Finally, as for the case $\Lambda \neq 0$, the components $(uu)$ and $(uA)$ of the Einstein equations yield time evolution constraints for the Bondi mass aspect $M$ and the angular momentum aspect $N^A$. 

In summary, the solution space for general relativity in the Bondi gauge with the preliminary boundary condition \eqref{asymp flat 1} is parametrized by the set of functions
\begin{equation}
\{q_{AB}, C_{AB}, M, N^A, \mathcal{D}_{AB}, \mathcal{E}_{AB}, \mathcal{F}_{AB}, \ldots \}_{\Lambda = 0} ,
\end{equation} where $q_{AB}$, $M$, $N^A$, $\mathcal{D}_{AB}$, $\mathcal{E}_{AB}$, $\mathcal{F}_{AB}$, $\ldots$ have constrained time evolutions (infinite tower of independent functions). Therefore, the characteristic initial value problem is well-defined when the following data are given: $C_{AB}(u,c^C)$, $q_{AB}(u_0, x^C)$, $M(u_0, x^C)$, $N^A(u_0, x^C)$, $\mathcal{D}_{AB}(u_0, x^C)$, $\mathcal{E}_{AB}(u_0, x^C)$, $\mathcal{F}_{AB}(u_0, x^C)$, $\ldots$ where $u_0$ is a fixed value of $u$. Notice a subtle point here: by taking the flat limit of the solution space with $\Lambda \neq 0$, we assumed that $g_{AB}$ is analytic in $r$ and can be expanded as \eqref{preliminary boundary condition} (this condition was not restrictive for $\Lambda \neq 0$). This condition is slightly more restrictive than \eqref{asymp flat 1} where analyticity is assumed only up to order $r^{-1}$. Therefore, by this flat limit procedure, we only obtain a subsector of the most general solution space. Writing $g_{AB}(u,r,x^C) = r^2 q_{AB}(u,x^C) + r C_{AB}(u,x^C) + D_{AB}(u,x^C) + \tilde{E}_{AB}  (u,r,x^C)$, where $\tilde{E}_{AB}$ is a function of all the coordinates of order $\mathcal{O}(r^{-1})$ in $r$, the most general solution space can be written as
\begin{equation}
\{q_{AB}, C_{AB}, M, N^A, \mathcal{D}_{AB}, \tilde{\mathcal{E}}_{AB} \}_{\Lambda = 0} ,
\end{equation} where $\tilde{\mathcal{E}}_{AB}$ is the trace-free part of $\tilde{{E}}_{AB}$, and $q_{AB}$, $M$, $N^A$, $\mathcal{D}_{AB}$, $\tilde{\mathcal{E}}_{AB}$ obey time evolution constraints. Now, the characteristic initial value problem is well-defined when the following data are given: $C_{AB}(u,c^C)$, $q_{AB}(u_0, x^C)$, $M(u_0, x^C)$, $N^A(u_0, x^C)$, $\mathcal{D}_{AB}(u_0, x^C)$ and $\tilde{\mathcal{E}}_{AB}(u_0,r, x^C)$. 

We complete this set of examples by mentioning the restricted solution spaces in the different definitions of asymptotic flatness introduced above. For boundary conditions (AF1) (equations \eqref{asymp flat 1} with \eqref{asymp flat 1 prime}), we obtain
\begin{equation}
\{q_{AB}, C_{AB}, M, N^A, \mathcal{D}_{AB}, \tilde{\mathcal{E}}_{AB} \}_{\Lambda = 0} ,
\end{equation} where $q_{AB}$, $M$, $N^A$, $\mathcal{D}_{AB}$ and $\tilde{\mathcal{E}}_{AB}$ obey time evolution constraints, and $\sqrt{q}$ is fixed. In particular, if we choose a branch where $\sqrt{q}$ is time-independent, from \eqref{time evolution constraint}, we immediately see that $\partial_u q_{AB} = 0$\footnote{The condition $\partial_u q_{AB} = 0$ was assumed for technical reasons in \cite{Campiglia:2014yka, Campiglia:2015yka, Compere:2018ylh}, but this was actually not restrictive.}. For boundary conditions (AF2) (equations \eqref{asymp flat 1} with \eqref{asymp flat 2}), the solution space reduces to 
\begin{equation}
\{\varphi, C_{AB}, M, N^A, \mathcal{D}_{AB}, \tilde{\mathcal{E}}_{AB} \}_{\Lambda = 0} ,
\end{equation} where $M$, $N^A$, $\mathcal{D}_{AB}$ and $\tilde{\mathcal{E}}_{AB}$ obey time evolution equations. Notice that the metric $q_{AB}$ of the form \eqref{asymp flat 2} automatically satisfies \eqref{time evolution constraint}. This agrees with results of \cite{Barnich:2010eb}. Finally, taking the boundary conditions (AF3) (equations \eqref{asymp flat 1} with \eqref{asymp flat 3}) yields the solution space
\begin{equation}
\{C_{AB}, M, N^A, \mathcal{D}_{AB}, \tilde{\mathcal{E}}_{AB} \}_{\Lambda = 0} ,
\label{solution space Bondi dirich}
\end{equation} where $M$, $N^A$, $\mathcal{D}_{AB}$ and $\tilde{\mathcal{E}}_{AB}$ obey time evolution equations. This agrees with the historical results of \cite{Bondi:1962px , Sachs:1962wk , Sachs:1962zza}.

\subsection{Asymptotic symmetry algebra}
\label{Asymptotic symmetry algebra}

\paragraph{Definition \normalfont[Asymptotic symmetry]} Given boundary conditions imposed in a chosen gauge, the \textit{asymptotic symmetries} are defined as the residual gauge transformations preserving the boundary conditions\footnote{This is the weak definition of asymptotic symmetry, in the sense of \eqref{ASG def 2}.}. In other words, the asymptotic symmetries considered on-shell are the gauge transformations $R[F]$ tangent to the solution space. In practice, the requirement to preserve the boundary conditions gives some constraints on the functions parametrizing the residual gauge transformations. In gravity, the generators of asymptotic symmetries are often called \textit{asymptotic Killing vectors}.

\paragraph{Definition \normalfont[Asymptotic symmetry algebra]} Once the asymptotic symmetries are known, we have
\begin{equation}
\begin{split}
[R[F_1], R[F_2]] &= \delta_{F_1} R[ F_2] - \delta_{F_2} R[ F_1] \\ 
&\approx R [[F_1, F_2]_A] ,
\end{split}
\label{representation on the solution space}
\end{equation} where $\approx$ means that this equality holds on-shell, i.e. on the solution space. In this expression, the bracket $[F_1, F_2]_A$ of gauge symmetry generators is given by
\begin{equation}
[F_1, F_2]_A = C(F_1, F_2) - \delta_{F_1} F_2 + \delta_{F_2} F_1 ,
\label{modified lie bracket}
\end{equation} where $C(F_1, F_2)$ is a skew-symmetric bi-differential operator \cite{Barnich:2017ubf , Barnich:2018gdh}
\begin{equation}
C(F_1, F_2) = \sum_{k,l \ge 0} C^{(\mu_1 \cdots \mu_k)(\nu_1 \cdots \nu_l)}_{[\alpha\beta]} \partial_{\mu_1} \ldots \partial_{\mu_k}f^\alpha_1 \partial_{\nu_1} \ldots \partial_{\nu_l} f_2^\beta .
\end{equation} The presence of the terms $- \delta_{F_1} F_2 + \delta_{F_2} F_1$ in \eqref{representation on the solution space} is due to the possible field-dependence of the asymptotic symmetry generators. We can verify that \eqref{modified lie bracket} satisfies the Jacobi identity, i.e. the asymptotic symmetry generators form a (solution space-dependent) Lie algebra for this bracket. It is called the \textit{asymptotic symmetry algebra}. The statement \eqref{representation on the solution space} means that the infinitesimal action of the gauge symmetries on the fields forms a representation of the Lie algebra of asymptotic symmetry generators: $[\delta_{F_1}, \delta_{F_2}] \Phi = \delta_{[F_1,F_2]_A} \Phi$. Let us mention that a Lie algebroid structure is showing up at this stage \cite{Crainic ,  Barnich:2010xq ,  Barnich:2017ubf}. The base manifold is given by the solution space, the field-dependent Lie algebra is the Lie algebra of asymptotic symmetry generators introduced above and the anchor is the map $F \to R[F]$.

\paragraph{Examples} Let us start by considering \textit{asymptotically AdS$_4$ spacetimes} in the \textit{Fefferman-Graham} and \textit{Bondi gauge}. The preliminary boundary condition \eqref{as ads} does not impose any constraint on the generators of the residual gauge diffeomorphisms of the Fefferman-Graham gauge given in \eqref{residual FG}. Similarly, the generators of the residual gauge diffeomorphisms in Bondi gauge given in \eqref{eq:xir} do not get further constraints with \eqref{as ads Bondi}. 

Now, let us consider the boundary conditions (AAdS1) (equation \eqref{as ads} together with \eqref{boundary gauge fixing FG}) in the Fefferman-Graham gauge. The asymptotic symmetries are generated by the vectors fields $\xi^\mu$ given in \eqref{residual FG} preserving the boundary conditions, namely, satisfying $\mathcal{L}_\xi g_{tt}^{(0)} =0$, $\mathcal{L}_\xi g_{tA}^{(0)} = 0$ and $g^{ab}_{(0)} \mathcal{L}_\xi g_{ab}^{(0)}= 0$. This leads to the following constraints on the parameters:
\begin{equation}
\left(\partial_u - \frac{1}{2} l  \right) \xi_0^t = \frac{1}{2} D_A^{(0)} \xi^A_0 , \quad \partial_u \xi^A_0  = - \frac{\Lambda}{3} g^{AB}_{(0)} \partial_B \xi_0^t, \quad \sigma = \frac{1}{2} (D_A^{(0)} \xi^A_0 + \xi^t_0 l )  ,
\label{constraint equations FG param}
\end{equation} where $l = \partial_u \ln \sqrt{\bar{q}}$. In this case, the Lie bracket \eqref{modified lie bracket} is given by 
\begin{equation}
[\xi_1, \xi_2]_A = \mathcal{L}_{\xi_1} \xi_2 - \delta_{\xi_1} \xi_2 + \delta_{\xi_2} \xi_1
\label{modified lie bracket gravity}
\end{equation}  and is referred as the \textit{modified Lie bracket} \cite{Barnich:2010eb}. Therefore, the asymptotic symmetry algebra can be worked out and is given explicitly by $[\xi (\xi_{0,1}^t, \xi^A_{0,1}), \xi (\xi_{0,2}^t, \xi^A_{0,2})]_A = \xi({\hat{\xi}}_0^t, {\hat{\xi}}_0^A)$, where
\begin{equation}
\begin{split}
{\hat{\xi}}_0^t &= \xi^A_{0,1} \partial_A \xi_{0,2}^t + \frac{1}{2} \xi_{0,1}^t  D_A^{(0)} \xi^A_{0,2} - \delta_{\xi (\xi_{0,1}^t, \xi^A_{0,1}) } \xi_{0,2}^t - (1 \leftrightarrow 2 ), \\
{\hat{\xi}}_0^A &= {\xi}_{0,1}^B \partial_B \xi^A_{0,2} - \frac{\Lambda}{3} \xi_{0,1}^t g^{AB}_{(0)} \partial_B \xi_{0,2}^t  - \delta_{\xi (\xi_{0,1}^t, \xi^A_{0,1}) } \xi_{0,2}^A - (1 \leftrightarrow 2 ) .\\
\end{split}
\label{ASG FG gen}
\end{equation} In the Bondi gauge with corresponding boundary conditions \eqref{boundary gauge fixing Bondi}, the constraints on the parameters are given by
\begin{equation}
\left(\partial_u - \frac{1}{2} l  \right) f = \frac{1}{2} D_A Y^A , \quad \partial_u Y^A  = - \frac{\Lambda}{3} q^{AB} \partial_B f , \quad \omega = 0
\label{constraint equations Bondi param}
\end{equation} and the asymptotic symmetry algebra is written as $[\xi (f_1, Y^A_{1}), \xi (f_{2}, Y^A_{2})]_A = \xi({\hat{f}}, {\hat{Y}}^A)$
where 
\begin{equation}
\begin{split}
{\hat{f}} &= Y^A_{1} \partial_A f_{2} + \frac{1}{2} f_{1}  D_A Y^A_2 - \delta_{\xi (f_1, Y^A_1 )} f_2  - (1 \leftrightarrow 2 ), \\
{\hat{Y}}^A &= {Y}_{1}^B \partial_B Y^A_2 - \frac{\Lambda}{3} f_{1} q^{AB} \partial_B f_{2} - \delta_{\xi (f_1, Y^A_1 )} Y^A_2 - (1 \leftrightarrow 2 ). \\
\end{split}
\label{ASG Bondi gen}
\end{equation} This asymptotic symmetry algebra is infinite-dimensional (in particular, it contains the area-preserving diffeomorphisms as a subgroup) and field-dependent, and it is called the $\Lambda$-BMS$_4$ algebra \cite{Compere:2019bua}. The parameters $f$ are called the supertranslation generators, while the parameters $Y^A$ are called the superrotation generators. These names will be justified below when studying the flat limit of this asymptotic symmetry algebra $\Lambda$-BMS$_4$. The computation of the modified Lie bracket \eqref{modified lie bracket gravity} in the Bondi gauge for these boundary conditions\footnote{This completes the results obtained in \cite{Compere:2019bua} where the asymptotic symmetry algebra was obtained by pullback methods.} follows closely \cite{Barnich:2010eb}. 

Let us consider the Fefferman-Graham gauge with Dirichlet boundary conditions\\(AAdS2), that is, \eqref{as ads} together with \eqref{BC Dirichlet}. Compared to the above situation, the equations \eqref{constraint equations FG param} reduce to
\begin{equation}
\partial_u  \xi_0^t = \frac{1}{2} D_A^{(0)} \xi^A_0 , \quad \partial_u Y^A  = - \frac{\Lambda}{3} \mathring{q}^{AB}_{(0)} \partial_B \xi_0^t, \quad \sigma = \frac{1}{2} D_A^{(0)} \xi^A_0  ,
\end{equation} where $D_A^{(0)}$ is the covariant derivative associated with the fixed unit sphere metric $\mathring{q}_{AB}$. Furthermore, there is an additional constraint: $\mathcal{L}_\xi g_{AB}^{(0)} = o(\rho^{-2})$, which indicates that $\xi^A_0$ is a conformal Killing vector of $\mathring{q}_{AB}$, namely,
\begin{equation}
D_A^{(0)} \xi^0_B + D_B^{(0)} \xi^0_A = D_C^{(0)} \xi^C_0 \mathring{q}_{AB} .
\end{equation} The asymptotic symmetry algebra remains of the same form as \eqref{ASG FG gen}. In the Bondi gauge, Dirichlet boundary conditions are given by \eqref{as ads Bondi} together with \eqref{BC Dirichlet Bondi}. The equations \eqref{constraint equations Bondi param} become
\begin{equation}
\partial_u  f = \frac{1}{2} D_A Y^A , \quad \partial_u Y^A  = - \frac{\Lambda}{3} \mathring{q}^{AB} \partial_B \xi_0^t , \quad \omega = 0 ,
\end{equation} where $D_{A}$ is the covariant derivative with respect to $\mathring{q}_{AB}$, while the additional constraint $\mathcal{L}_\xi g_{AB} = o(r^2)$ gives
\begin{equation}
D_A Y_B + D_B Y_A = D_C Y^C \mathring{q}_{AB} .
\end{equation} This means that $Y^A$ is a conformal Killing vector of $\mathring{q}_{AB}$. The asymptotic symmetry algebra \eqref{ASG Bondi gen} remains of the same form. It can be shown that the asymptotic symmetry algebra corresponds to $SO(3,2)$ algebra for $\Lambda<0$ and $SO(1,4)$ algebra for $\Lambda>0$ \cite{Barnich:2013sxa} (see also Appendix A of \cite{Compere:2019bua}). Therefore, we see how the infinite-dimensional asymptotic symmetry algebra $\Lambda$-BMS$_4$ reduces to these finite-dimensional algebras, which are the symmetry algebras of global AdS$_4$ and global dS$_4$, respectively.  

Let us now consider \textit{four-dimensional asymptotically flat spacetimes} in the \textit{Bondi gauge}. The asymptotic Killing vectors can be derived in a similar way to that in the previous examples. Another way in which to proceed is to take the flat limit of the previous results obtained in the Bondi gauge. We sketch the expressions obtained by following these two equivalent procedures. First, consider the preliminary boundary conditions \eqref{asymp flat 1}. The \textit{asymptotic Killing vectors} $\xi^\mu$ are the residual gauge diffeomorphisms \eqref{eq:xir} with the following constraints on the parameters:
\begin{equation}
\left(\partial_u - \frac{1}{2} l  \right) f = \frac{1}{2} D_A Y^A-  \omega , \quad \partial_u Y^A  = 0  ,
\end{equation} where $l =\partial_u \ln \sqrt{q}$. These equations can be readily solved and the solutions are given by
\begin{equation}
f = q^{\frac{1}{4}} \left[ T(x^A) + \frac{1}{2} \int^u_0 \mathrm{d}u' [ q^{-\frac{1}{4}} (D_A Y^A - 2 \omega) ] \right], \quad Y^A = Y^A(x^B) ,
\end{equation} where $T$ are called supertranslation generators and $Y^A$ superrotation generators. Notice that there is no additional constraint on $Y^A$ at this stage. Computing the modified Lie bracket \eqref{modified lie bracket gravity}, we obtain $[\xi (f_1, Y^A_{1},\omega_1), \xi (f_{2}, Y^A_{2}, \omega_2)]_A = \xi({\hat{f}}, {\hat{Y}}^A, \hat{\omega})$ where
\begin{equation}
\begin{split}
{\hat{f}} &= Y^A_{1} \partial_A f_{2} + \frac{1}{2} f_{1}  D_A Y^A_2 - (1 \leftrightarrow 2 ) , \\
{\hat{Y}}^A &= {Y}_{1}^B \partial_B Y^A_2 - (1 \leftrightarrow 2 ), \\
\hat{\omega} &= 0 .
\end{split}
\label{asg flat}
\end{equation} 

Now, we discuss the two relevant sub-cases of boundary conditions in asymptotically flat spacetimes. Adding the condition \eqref{asymp flat 1 prime} to the preliminary condition \eqref{asymp flat 1}, i.e. considering (AF1), gives the additional constraint 
\begin{equation}
\omega = 0
\end{equation} Note that this case corresponds exactly to the flat limit of the (AAdS1) case (equations \eqref{as ads} and \eqref{boundary gauge fixing FG}). The asymptotic symmetry algebra reduces to the semi-direct product
\begin{equation}
\text{Diff($S^2$)} \ltimes \mathcal{S} ,
\label{generalized BMS group}
\end{equation} where $\text{Diff($S^2$)}$ are the smooth superrotations generated by $Y^A$ and $\mathcal{S}$ are the smooth supertranslations generated by $T$. This extension of the original global BMS$_4$ algebra (see below) is called the \textit{generalized BMS$_4$ algebra} \cite{Campiglia:2014yka , Campiglia:2015yka , Compere:2018ylh , Flanagan:2019vbl}. Therefore, the $\Lambda$-BMS$_4$ algebra reduces in the flat limit to the smooth extension \eqref{generalized BMS group} of the BMS$_4$ algebra.  

The other sub-case of boundary conditions for asymptotically flat spacetimes (AF2) is given by adding condition \eqref{asymp flat 2} to the preliminary boundary condition \eqref{asymp flat 1}. The additional constraint on the parameters is now given by
\begin{equation}
D_A Y_B + D_B Y_A = D_C Y^C \mathring{q}_{AB} ,
\end{equation} i.e. $Y^A$ is a conformal Killing vector of the unit round sphere metric $\mathring{q}_{AB}$. If we allow $Y^A$ to not be globally well-defined on the $2$-sphere, then the asymptotic symmetry algebra has the structure
\begin{equation}
(\text{Vir}\times\text{Vir} \ltimes \mathcal{S}^*) \times \mathbb{R} .
\end{equation}  Here, $\text{Vir}\times\text{Vir}$ is the direct product of two copies of the Witt algebra, parametrized by $Y^A$. Furthermore, $\mathcal{S}^*$ are the supertranslations, parametrized by $T$, and $\mathbb{R}$ are the abelian Weyl rescalings of $\mathring{q}_{AB}$, parametrized by $\omega$. Note that the supetranslations also contain singular elements since they are related to the singular superrotations through the algebra \eqref{asg flat}. This extension of the global BMS$_4$ algebra is called the \textit{extended BMS$_4$ algebra} \cite{Barnich:2010eb}. Finally, as a sub-case of this one, considering the more restrictive constraints \eqref{asymp flat 3}, i.e. (AF3), and allowing only globally well-defined $Y^A$, we recover the \textit{global BMS$_4$ algebra} \cite{Bondi:1962px , Sachs:1962wk , Sachs:1962zza}, which is given by
\begin{equation}
SO(3,1) \ltimes \mathcal{S} ,
\end{equation} where $\mathcal{S}$ are the supertranslations and $SO(3,1)$ is the algebra of the globally well-defined conformal Killing vectors of the unit $2$-sphere metric, which is isomorphic to the proper orthocronous Lorentz group in four dimensions. 

\paragraph{Definition \normalfont[Action on the solution space]} Given boundary conditions imposed in a chosen gauge, there is a natural \textit{action of the asymptotic symmetry algebra}, with generators $F=F(a)$, \textit{on the solution space} $\tilde{\Phi}= \tilde{\Phi}(b)$. The form of this action can be deduced from \eqref{transformation of the fields} by inserting the solution space and the explicit form of the asymptotic symmetry generators\footnote{This action is usually not linear. However, in three-dimensional general relativity, this action is precisely the coadjoint representation of the asymptotic symmetry algebra \cite{Barnich:2012rz, Barnich:2014zoa , Barnich:2015uva , Barnich:2017jgw , Oblak:2017ect}.}.

\paragraph{Examples} In the \textit{Fefferman-Graham gauge} with Dirichlet boundary conditions for \textit{asymptotically AdS$_4$ spacetimes} (AAdS2) (\eqref{as ads} with \eqref{BC Dirichlet}), the asymptotic symmetry algebra $SO(3,2)$ acts on the solution space \eqref{solution space dirichlet} as 
\begin{equation}
\delta_{\xi^c_0} T_{ab} = \left( \mathcal{L}_{\xi^c_0} + \frac{1}{3} D^{(0)}_c \xi^c_0 \right) T_{ab}.
\end{equation} In the \textit{Bondi gauge} with definition (AF3) (\eqref{asymp flat 1} with \eqref{asymp flat 3}) of \textit{asymptotically flat spacetime}, the global BMS$_4$ algebra acts on the leading functions of the solution space \eqref{solution space Bondi dirich} as 
\begin{equation}
\begin{split}
\delta_{(f,Y)} C_{AB} &= [f \partial_u + \mathcal{L}_Y - \frac{1}{2} D_C Y^C ] C_{AB} - 2 D_A D_B f + \mathring{q}_{AB} D_C D^C f,\\
\delta_{(f,Y)} M &= [f \partial_u + \mathcal{L}_Y + \frac{3}{2} D_C Y^C] M  + \frac{1}{4} N^{AB} D_A D_B f + \frac{1}{2} D_A f D_B N^{AB} + \frac{1}{8} D_C D_B D_A Y^A C^{BC},\\
\delta_{(f,Y)} N_A &= [f\partial_u + \mathcal{L}_Y + D_C Y^C] N_A + 3 M D_A f - \frac{3}{16} D_A f N_{BC} C^{BC} + \frac{1}{2} D_B f N^{BC} C_{AC} \nonumber \\
&\quad - \frac{1}{32} D_A D_B Y^B C_{CD}C^{CD} + \frac{1}{4} (2 D^B f  + D^B D_C D^C f) C_{AB} \nonumber \\
&\quad - \frac{3}{4} D_B f (D^B D^C C_{AC} - D_A D_C C^{BC}) + \frac{3}{8} D_A (D_C D_B f C^{BC}) \nonumber \\
&\quad + \frac{1}{2} (D_A D_B f - \frac{1}{2} D_C D^C f \mathring{q}_{AB}) D_C C^{BC},
\end{split}
\end{equation} where $N_{AB} = \partial_u C_{AB}$ \cite{Barnich:2010eb}. For the action of the associated asymptotic symmetry group on these solution spaces, see \cite{Barnich:2016lyg}.

\section{Surface charges}
\label{Surface charges}

In this section, we review how to construct the surface charges associated with gauge symmetries. After recalling some results about global symmetries and Noether currents, the Barnich-Brandt prescription to obtain the surface charges in the context of asymptotic symmetries is discussed. We illustrate this construction with the example of general relativity with asymptotically (A)dS and asymptotically flat spacetimes. The relation between this prescription and the covariant phase space methods is established.

\subsection{Global symmetries and Noether's first theorem}

\paragraph{Definition \normalfont[Global symmetry]} Let us consider a Lagrangian theory with Lagrangian density $\mathbf{L}[\Phi, \partial_\mu \Phi, \ldots]$ and a transformation $\delta_Q \Phi = Q$ of the fields, where $Q$ is a local function. In agreement with the above definition \eqref{symmetry}, this transformation is said to be a \textit{symmetry} of the theory if
\begin{equation}
\delta_Q \mathbf{L} = \mathrm{d} \mathbf{B}_Q ,
\label{symmetry def}
\end{equation} where $\mathbf{B}_Q = B^\mu_Q (\mathrm{d}^{n-1}x)_\mu$. Then, as defined in \eqref{transformation of the fields}, a \textit{gauge symmetry} is just a symmetry that depends on arbitrary spacetime functions $F = (f^\alpha)$, i.e. $Q = R[F]$. We define an on-shell equivalence relation $\sim$ between the symmetries of the theory as
\begin{equation}
Q \sim Q + R[F] ,
\end{equation} i.e. two symmetries are equivalent if they differ, on-shell, by a gauge transformation $R[F]$. The equivalence classes $[Q]$ for this equivalence relation are called the \textit{global symmetries}. In particular, a gauge symmetry is a trivial global symmetry. 

\paragraph{Definition \normalfont[Noether current]} A \textit{conserved current} $\mathbf{j}$ is an on-shell closed $(n-1)$-form, i.e. $\mathrm{d} \mathbf{j} \approx 0$. We define an on-shell equivalence relation $\sim$ between the currents as
\begin{equation}
\mathbf{j} \sim \mathbf{j} + \mathrm{d} \mathbf{K} ,
\label{Noether current def}
\end{equation} where $\mathbf{K}$ is a $(n-2)$-form. A \textit{Noether current} is an equivalence class $[\mathbf{j}]$ for this equivalence relation.

\paragraph{Theorem \normalfont[Noether's first theorem]} A one-to-one correspondence exists between \textit{global symmetries} $Q$ and \textit{Noether currents} $[\mathbf{j}]$, which can be written as
\begin{equation}
[Q] \stackrel{\text{\scriptsize 1-1}}{\longleftrightarrow} [\mathbf{j}] .
\label{one to one}
\end{equation} In particular, Noether currents associated with gauge symmetries are trivial. Recent demonstrations of this theorem can for example be found in \cite{Barnich:2018gdh , Barnich:2001jy}. 

\paragraph{Remark} This theorem also enables us to construct explicit representatives of the Noether current for a given global symmetry. We have
\begin{equation}
\delta_Q \mathbf{L} = \mathrm{d}\mathbf{B}_Q = (\partial_\mu B^\mu_Q) \mathrm{d}^n x .
\label{stage 1}
\end{equation} Furthermore, writing $\mathbf{L} = L \mathrm{d}^n x$, we obtain 
\begin{equation}
\begin{split}
\delta_Q L &= \delta_Q \Phi \frac{\partial L}{\partial  \Phi} + \delta_Q \partial_\mu \Phi \frac{\partial L}{\partial (\partial_\mu \Phi)} + \ldots \\
&= Q \frac{\partial L}{\partial \Phi} + \partial_\mu Q \frac{\partial L}{\partial (\partial_\mu \Phi)} + \ldots \\
&= Q \left( \frac{\partial L}{\partial \Phi} - \partial_\mu \frac{\partial L}{\partial (\partial_\mu \Phi)} + \ldots \right) + \partial_\mu \left( Q \frac{\partial L}{\partial (\partial_\mu \Phi)} + \ldots \right) \\
&= Q \frac{\delta L}{\delta \Phi} + \partial_\mu \left( Q \frac{\partial L}{\partial (\partial_\mu \Phi)} + \ldots \right) ,
\end{split}
\label{stage 2}
\end{equation} where, in the second line, we used 
\begin{equation}
[\delta_Q, \partial_\mu] = 0 
\label{commutator}
\end{equation} and, in the last equality, we used \eqref{Euler lagrange def}. Putting \eqref{stage 1} and \eqref{stage 2} together, we obtain
\begin{equation}
Q \frac{\delta L}{\delta \Phi} = \partial_\mu \left( B^\mu_Q - Q \frac{\partial L}{\partial (\partial_\mu \Phi)} + \ldots  \right) \equiv \partial_\mu j^\mu_Q
\end{equation} or, equivalently
\begin{equation}
Q \frac{\delta \mathbf{L}}{\delta \Phi} = \mathrm{d} \mathbf{j}_Q ,
\label{equality important}
\end{equation}  where $\mathbf{j}_Q = j^\mu_Q (\mathrm{d}^{n-1}x)_\mu$. In particular, $\mathrm{d} \mathbf{j}_Q \approx 0$ holds on-shell. Hence, we have obtained a representative of the Noether current associated with the global symmetry $Q$ through the correspondence \eqref{one to one}. 

\paragraph{Theorem \normalfont[Noether representation theorem]} Defining the bracket as
\begin{equation}
\{ \mathbf{j}_{Q_1} , \mathbf{j}_{Q_2} \} = \delta_{Q_1} \mathbf{j}_{Q_2} ,
\end{equation} we have
\begin{equation}
\{ \mathbf{j}_{Q_1} , \mathbf{j}_{Q_2} \} \approx \mathbf{j}_{[Q_1, Q_2]} 
\label{algebra current Noether}
\end{equation} ($n>1$), where $[Q_1,Q_2] = \delta_{Q_1} Q_2 - \delta_{Q_2} Q_1$. In other words, the Noether currents form a representation of the symmetries. 

To prove this theorem, we apply $\delta_{Q_1}$ on the left-hand side and the right-hand side of \eqref{equality important}, where $Q$ is replaced by $Q_2$. On the right-hand side, using the first equation of \eqref{commutation relation delta Q}, we obtain
\begin{equation}
\delta_{Q_1}\mathrm{d} \mathbf{j}_{Q_2} \approx \mathrm{d} \delta_{Q_1} \mathbf{j}_{Q_2} .
\label{equa 1}
\end{equation} On the left-hand side, we have
\begin{equation}
\begin{split}
\delta_{Q_1} \left( Q_2 \frac{\delta \mathbf{L}}{\delta \Phi} \right) &= \delta_{Q_1} Q_2 \frac{\delta \mathbf{L}}{\delta \Phi} + Q_2 \delta_{Q_1} \frac{\delta \mathbf{L}}{\delta \Phi}  \\
&=\delta_{Q_1} Q_2 \frac{\delta \mathbf{L}}{\delta \Phi} +  Q_2 \frac{\delta }{\delta \Phi} (\delta_{Q_1} \mathbf{L} ) -  Q_2 \sum_{k \ge 0} (-1)^k \partial_{\mu_1} \ldots \partial_{\mu_k} \left( \frac{\partial Q_1}{\partial \Phi_{\mu_1\ldots\mu_k}} \frac{\delta \mathbf{L}}{\delta \Phi} \right) \\
&= \delta_{Q_1} Q_2 \frac{\delta \mathbf{L}}{\delta \Phi}-  Q_2 \sum_{k \ge 0} (-1)^k \partial_{\mu_1} \ldots \partial_{\mu_k} \left( \frac{\partial Q_1}{\partial \Phi_{\mu_1\ldots\mu_k}} \frac{\delta \mathbf{L}}{\delta \Phi} \right) , \\
\end{split}
\end{equation} where, to obtain the second equality, we used \eqref{variation and euler lagrange}. In the last equality, we used \eqref{symmetry def} together with \eqref{euler lagrange and total}. Now, using Leibniz rules in the second term of the right-hand side, we get 
\begin{equation}
\begin{split}
\delta_{Q_1} \left( Q_2 \frac{\delta \mathbf{L}}{\delta \Phi} \right) &= \delta_{Q_1} Q_2 \frac{\delta \mathbf{L}}{\delta \Phi}- \sum_{k \ge 0}  \partial_{\mu_1} \ldots \partial_{\mu_k} Q_2  \left( \frac{\partial Q_1}{\partial \Phi_{\mu_1\ldots\mu_k}} \frac{\delta \mathbf{L}}{\delta \Phi} \right) + \partial_\mu T^\mu_{Q_1} \left( Q_2,  \frac{\delta L}{\delta \Phi} \right) \mathrm{d}^n x \\
&= ( \delta_{Q_1} Q_2 - \delta_{Q_2} Q_1 ) \frac{\delta \mathbf{L}}{\delta \Phi} + \partial_\mu T^\mu_{Q_1} \left( Q_2,  \frac{\delta L}{\delta \Phi} \right)  \mathrm{d}^n x \\
&= [Q_1, Q_2]  \frac{\delta \mathbf{L}}{\delta \Phi} + \partial_\mu T^\mu_{Q_1} \left( Q_2,  \frac{\delta L}{\delta \Phi} \right) d^n x \\
&= \mathrm{d} \mathbf{j}_{[Q_1, Q_2]} + \mathrm{d} \mathbf{T}_{Q_1} \left( Q_2,  \frac{\delta L}{\delta \Phi} \right) ,
\end{split}
\label{equa 2}
\end{equation} where $T^\mu_{Q_1} \left( Q_2,  \frac{\delta L}{\delta \Phi} \right)$ is an expression vanishing on-shell. In the second equality, we used \eqref{def variation}, and in the last equality, we used \eqref{equality important}. Putting \eqref{equa 1} and \eqref{equa 2} together results in 
\begin{equation}
\mathrm{d} \left[ \delta_{Q_1} \mathbf{j}_{Q_2} - \mathbf{j}_{[Q_1, Q_2]} - \mathbf{T}_{Q_1} \left( Q_2,  \frac{\delta L}{\delta \Phi} \right) \right] = 0 .
\label{closed}
\end{equation} We know from Poincar\'e lemma that locally, every closed form is exact\footnote{The Poincar\'e lemma states that in a star-shaped open subset, the de Rham cohomology class $H^p_{dR}$ is given by \[H^p_{dR} = \left\{
    \begin{array}{ll}
         0 & \mbox{if } 0<p \le n \\
        \mathbb{R} & \mbox{if } p =0
    \end{array} .
\right.\]}. However, this cannot be the case in Lagrangian field theories. In fact, this would imply that every $n$-form is exact, and therefore, there would not be any possibility of non-trivial dynamics. Let us remark that the operator $\mathrm{d}$ that we are using is not the standard exterior derivative, but a horizontal derivative in the jet bundle (see definition \eqref{horizontal}) that takes into account the field-dependence. In this context, we have to use the \textit{algebraic Poincar\'e lemma}.

\paragraph{Lemma \normalfont[Algebraic Poincar\'e lemma]} The cohomology class $H^p(\mathrm{d})$ for the operator $\mathrm{d}$ defined in \eqref{horizontal} is given by
\begin{equation}
H^p(\mathrm{d}) = \left\{
    \begin{array}{lll}
        [\boldsymbol{\alpha}^n] & \mbox{if } p = n \\
         0 & \mbox{if } 0<p<n \\
        \mathbb{R} & \mbox{if } p =0
    \end{array}
\right.
\label{Poincare lemma}
\end{equation} where $[\boldsymbol{\alpha}^n]$ designates the equivalence classes of $n$-forms for the equivalence relation $\boldsymbol{\alpha}^n \sim {\boldsymbol{\alpha}'}^n$ if $\boldsymbol{\alpha}^n = {\boldsymbol{\alpha}'}^n + \mathrm{d} \boldsymbol{\beta}^{n-1}$ \cite{Barnich:2018gdh}.

Le us go back to the proof of \eqref{algebra current Noether}. Applying the algebraic Poincar\'e lemma to \eqref{closed} yields 
\begin{equation}
\delta_{Q_1} \mathbf{j}_{Q_2} = \mathbf{j}_{[Q_1, Q_2]} + \mathbf{T}_{Q_1} \left( Q_2,  \frac{\delta L}{\delta \Phi} \right) + \mathrm{d}\boldsymbol{\eta} ,
\end{equation} where $\boldsymbol{\eta}$ is a $(n-2)$-form.  Therefore, on-shell, since $\mathbf{T}_{Q_1} \left( Q_2,  \frac{\delta L}{\delta \Phi} \right) \approx 0$ and because Noether currents are defined up to exact $(n-1)$-forms, we obtain the result \eqref{algebra current Noether}. Notice that in classical mechanics (i.e. $n=1$), from \eqref{Poincare lemma}, constant central extensions may appear in the current algebra.

\paragraph{Definition \normalfont[Noether charge]} Given a Noether current $[\mathbf{j}]$, we can construct a \textit{Noether charge} by integrating it on a $(n-1)$-dimensional spacelike surface $\Sigma$, with boundary $\partial \Sigma$, as
\begin{equation}
H_Q[\Phi] = \int_\Sigma \mathbf{j} .
\label{Noether charge}
\end{equation} If we assume that the currents and their ambiguities vanish at infinity, this definition does not depend on the representative of the Noether current. Indeed, 
\begin{equation}
H'_Q[\Phi] = \int_\Sigma (\mathbf{j} + \mathrm{d} \mathbf{K}) = H_Q[\Phi] + \int_{\partial \Sigma} \mathbf{K} ,
\end{equation} where we used the Stokes theorem. Since $\int_{\partial \Sigma} \mathbf{K} =0$, we have $H'_Q[\Phi] = H_Q[\Phi]$.

\paragraph{Remark \normalfont[Conservation and algebra of Noether charges]} The Noether charge \eqref{Noether charge} is conserved in time, that is,
\begin{equation}
\frac{d}{dt} H_Q[\Phi] \approx 0 .
\end{equation} In fact, consider two spacelike hypersurfaces $\Sigma_1 \equiv t_1 = 0$ and $\Sigma_2 \equiv t_2 = 0$. We have
\begin{equation}
H^{t_2}_Q[\Phi] - H^{t_1}_Q[\Phi] = \int_{\Sigma_2} \mathbf{j}_Q  - \int_{\Sigma_1} \mathbf{j}_Q = \int_{\Sigma_2 - \Sigma_1} \mathrm{d} \mathbf{j}_Q \approx 0 ,
\end{equation} where $\Sigma_2 - \Sigma_1$ is the spacetime volume encompassed between $\Sigma_1$ and $\Sigma_2$. In the second equality, we used the hypothesis that currents vanish at infinity and the Stokes theorem. 

The Noether charges \eqref{Noether charge} form a representation of the algebra of global symmetries, i.e.
\begin{equation}
\{ H_{Q_1}, H_{Q_2} \} \approx H_{[Q_1, Q_2]} ,
\end{equation} where the bracket of Noether charges is defined as 
\begin{equation}
\{ H_{Q_1}, H_{Q_2} \} = \delta_{Q_1} H_{Q_2} = \int_\Sigma \delta_{Q_1} \mathbf{j}_{Q_2} .
\end{equation} This is a direct consequence of \eqref{algebra current Noether}.

\subsection{Gauge symmetries and lower degree conservation law}

\paragraph{Definition \normalfont[Noether identities]} Consider the relation \eqref{equality important} for a gauge symmetry:
\begin{equation}
R[F] \frac{\delta L}{\delta \Phi} = \partial_\mu j^\mu_F .
\label{equation importante gauge}
\end{equation} The left-hand side can be worked out as
\begin{equation}
\begin{split}
R[F] \frac{\delta L}{\delta \Phi} =& \left( R_\alpha f^\alpha + R^\mu_\alpha  \partial_\mu f^\alpha + R^{(\mu\nu)}_\alpha \partial_\mu \partial_\nu f^\alpha + \ldots \right) \frac{\delta L}{\delta \Phi} \\
=& f^\alpha \left[ R_\alpha  \frac{\delta L}{\delta \Phi} - \partial_\mu \left( R^\mu_\alpha   \frac{\delta L}{\delta \Phi} \right) +  \partial_\mu \partial_\nu \left( R^{(\mu\nu)}_\alpha \frac{\delta L}{\delta \Phi} \right) + \ldots \right] \\
&+ \partial_\mu \underbrace{\left[ R^\mu_\alpha f^\alpha  \frac{\delta L}{\delta \Phi} - f^\alpha \partial_\nu \left( R^{(\mu\nu)}_\alpha \frac{\delta L}{\delta \Phi}  \right) + \ldots \right]}_{\equiv S^\mu_F} .
\end{split}
\label{Noether identity proof}
\end{equation} Therefore, the equation \eqref{equation importante gauge} can be rewritten as
\begin{equation}
f^\alpha R^\dagger_\alpha \left( \frac{\delta L}{\delta \Phi}  \right) = \partial_\mu (j^\mu_F - S^\mu_F )  ,
\label{Noether proof 2}
\end{equation} where $R^\dagger_\alpha \left( \frac{\delta L}{\delta \Phi}  \right) = R_\alpha  \frac{\delta L}{\delta \Phi} - \partial_\mu \left( R^\mu_\alpha   \frac{\delta L}{\delta \Phi} \right) +  \partial_\mu \partial_\nu \left( R^{(\mu\nu)}_\alpha \frac{\delta L}{\delta \Phi} \right) + \ldots$ Since $F$ is a set of arbitrary functions, we can apply the Euler-Lagrange derivative \eqref{Euler lagrange def} with respect to $f^\alpha$ on this equation. Since the right-hand side is a total derivative, it vanishes under the action of the Euler-Lagrange derivative (see \eqref{euler lagrange and total}) and we obtain
\begin{equation}
R^\dagger_\alpha \left( \frac{\delta \mathbf{L}}{\delta \Phi}  \right) = 0 .
\label{Noether identity}
\end{equation} This identity is called a \textit{Noether identity}. There is one identity for each independent generator $f^\alpha$. Notice that these identities are satisfied off-shell.

\paragraph{Theorem \normalfont[Noether's second theorem]} We have 
\begin{equation}
R[F] \frac{\delta \mathbf{L}}{\delta \Phi} = \mathrm{d} \mathbf{S}_F \left[\frac{\delta L}{\delta \Phi} \right] ,
\label{second Noether theorem}
\end{equation} where $\mathbf{S}_F = S^\mu_F (\mathrm{d}^{n-1} x)_\mu$ is the weakly vanishing Noether current (i.e. $\mathbf{S}_F \approx 0$) that was defined in the last line of \eqref{Noether identity proof}. This is a direct consequence of \eqref{Noether identity proof}, taking the Noether identity \eqref{Noether identity} into account.

\paragraph{Example} Consider the theory of general relativity $\mathbf{L} = (16\pi G)^{-1} (R - 2 \Lambda) \sqrt{-g} \mathrm{d}^n x$. The Euler-Lagrange derivative of the Lagrangian is given by 
\begin{equation}
\frac{\delta \mathbf{L}}{\delta g_{\mu\nu}} = -(16\pi G)^{-1}(G^{\mu\nu} + g^{\mu\nu} \Lambda)   \sqrt{-g} \mathrm{d}^n x .
\end{equation} The Noether identity associated with the diffeomorphism generated by $\xi^\mu$ is obtained by following the lines of \eqref{Noether identity proof}:
\begin{equation}
\begin{split}
(16 \pi G)\delta_\xi g_{\mu\nu} \frac{\delta \mathbf{L}}{\delta g_{\mu\nu}} &= -2 \nabla_\mu \xi_\nu (G^{\mu\nu} + g^{\mu\nu} \Lambda)   \sqrt{-g} \mathrm{d}^n x \\
&= 2\xi_\nu \nabla_\mu G^{\mu\nu} \sqrt{-g} \mathrm{d}^n x - \partial_\mu [ 2 \xi_\nu (G^{\mu\nu} + g^{\mu\nu} \Lambda)\sqrt{-g}] \mathrm{d}^n x .
\end{split}
\end{equation} Therefore, the Noether identity is the Bianchi identity for the Einstein tensor
\begin{equation}
\nabla_\mu G^{\mu \nu} = 0
\end{equation} and the weakly vanishing Noether current of Noether's second theorem \eqref{second Noether theorem} is given by
\begin{equation}
\mathbf{S}_\xi = -\frac{\sqrt{-g}}{8\pi G}\xi_\nu (G^{\mu\nu} + g^{\mu\nu} \Lambda) (\mathrm{d}^{n-1}x)_\mu .
\label{weakly vanishin noether current gravity}
\end{equation}

\paragraph{Remark} From \eqref{equation importante gauge} and \eqref{second Noether theorem}, we have $\mathrm{d} (\mathbf{j}_F - \mathbf{S}_F ) = 0$, and hence, from the algebraic Poincar\'e lemma \eqref{Poincare lemma},
\begin{equation}
\mathbf{j}_F =  \mathbf{S}_F + \mathrm{d} \mathbf{K}_F ,
\label{ambiguity}
\end{equation} where $\mathbf{K}_F$ is a $(n-2)$-form. Therefore, as already stated in Noether's first theorem \eqref{one to one}, the Noether current associated with a gauge symmetry is trivial, i.e. vanishing on-shell, up to an exact $(n-1)$-form. A natural question arises at this stage: is it possible to define a notion of conserved quantity for gauge symmetries? Naively, following the definition \eqref{Noether charge}, one may propose the following definition for conserved charge:
\begin{equation}
H_F = \int_\Sigma \mathbf{j}_F \approx \int_{\partial \Sigma} \mathbf{K}_F
\label{bad noether charge}
\end{equation} where, in the second equality, we used \eqref{ambiguity} and Stokes' theorem. This charge will be conserved on-shell since $\mathrm{d}\mathbf{j}_F \approx 0$. The problem is that the $(n-2)$-form $\mathbf{K}_F$ appearing in \eqref{bad noether charge} is completely arbitrary. Indeed, the Noether currents are equivalence classes of currents (see equation \eqref{Noether current def}). Therefore, we have to find an appropriate procedure to isolate a particular $\mathbf{K}_F$. 

%As we will see in the following, several procedures exist leading to different choices of $\mathbf{k}_F$. 

\paragraph{Definition \normalfont[Reducibility parameter]} \textit{Reducibility parameters} $\bar{F}$ are parameters of gauge transformations satisfying
\begin{equation}
R[\bar{F}] \approx 0 .
\end{equation} Two reducibility parameters $\bar{F}$ and $\bar{F}'$ are said to be equivalent, i.e. $\bar{F}\sim \bar{F}'$, if $\bar{F} \approx \bar{F}'$. Note that for a large class of gauge theories (including electrodynamics, Yang-Mills and general relativity in dimensions superior or equal to three \cite{Barnich:2001jy , Barnich:2018gdh}), these equivalence classes of asymptotic reducibility parameters are determined by field-independent ordinary functions $\bar{F}(x)$ satisfying the off-shell condition
\begin{equation}
R[\bar{F}] = 0 .
\label{exact red param}
\end{equation} We will call them \textit{exact reducibility parameters}.

\paragraph{Theorem \normalfont[Generalized Noether's theorem]} A one-to-one correspondence exists between \textit{equivalence classes of reducibility parameters} and \textit{equivalence classes of on-shell conserved $(n-2)$-forms} $[\mathbf{K}]$, which can be written as
\begin{equation}
[\bar{F}] \stackrel{\text{\scriptsize 1-1}}{\longleftrightarrow} [\mathbf{K}] .
\label{one to one gen}
\end{equation} In this statement, two conserved $(n-2)$-forms $\mathbf{K}$ and $\mathbf{K}'$ are said to be equivalent, i.e. $\mathbf{K}\sim \mathbf{K}'$, if $\mathbf{K} \approx \mathbf{K}' + \mathrm{d}\mathbf{l}$ where $\mathbf{l}$ is a $(n-3)$-form \cite{Barnich:1994db, Barnich:1995ap}. 

\paragraph{Remark} The Barnich-Brandt procedure allows for the construction of explicit representatives of the conserved $(n-2)$-forms for given exact reducibility parameters $\bar{F}$ \cite{Barnich:2001jy , Barnich:2003xg}. From Noether's second theorem \eqref{second Noether theorem} and \eqref{exact red param}, we have
\begin{equation}
\mathrm{d} \mathbf{S}_{\bar{F}} = 0 .
\label{weakly vanishing is conserved}
\end{equation} From the algebraic Poincar\'e Lemma \eqref{Poincare lemma}, we get\footnote{The minus sign on the left-hand side of \eqref{weakly vanishing is conserved conseq} is a matter of convention.}
\begin{equation}
-\mathrm{d} \mathbf{K}_{\bar{F}} = \mathbf{S}_{\bar{F}} \approx 0 .
\label{weakly vanishing is conserved conseq}
\end{equation} Using the homotopy operator \eqref{homotopy operator}, we define 
\begin{equation}
\mathbf{k}_{\bar{F}}[\Phi; \delta \Phi] = - I_{\delta \Phi}^{n-1} \mathbf{S}_{\bar{F}} .
\label{n-2 in case reduc}
\end{equation} This $\mathbf{k}_{\bar{F}}[\Phi; \delta \Phi]$ is an element of $\Omega^{n-2,1}$ (see appendix \ref{Useful results}) and is defined up to an exact $(n-2)$-form. This enables us to find an explicit expression for the conserved $(n-2)$-form $\mathbf{K}_{\bar{F}}[\Phi]$ as
\begin{equation}
\mathbf{K}_{\bar{F}}[\Phi] = \int_\gamma \mathbf{k}_{\bar{F}} [\Phi; \delta \Phi] ,
\label{DEF OF K}
\end{equation} where $\gamma$ is a path on the solution space relating $\bar{\Phi}$ such that $S_{\bar{F}}[\bar{\Phi}] = 0$ to the solution $\Phi$ of interest. Applying the operator $\mathrm{d}$ on \eqref{DEF OF K} gives back \eqref{weakly vanishing is conserved conseq}, using the property \eqref{commutation homotopy op} of the homotopy operator. Notice that the expression \eqref{DEF OF K} of $\mathbf{K}_{\bar{F}}[\Phi]$ generically depends on the chosen path $\gamma$. Therefore, in practice, we consider the $(n-2)$-form $\mathbf{k}_{\bar{F}}[\Phi; \delta \Phi]$ defined in \eqref{n-2 in case reduc} as the fundamental object, rather than $\mathbf{K}_{\bar{F}}[\Phi]$.  

%--------------------------------
%
%\begin{equation}
%\begin{split}
%\mathbf{K}_{\bar{F}}[\Phi] &= \int_\gamma \mathbf{k}_F [\Phi, \delta \Phi] \\
%&= \int_\gamma - I_{\delta \Phi}^{n-1} \mathbf{S}_{\bar{F}} \\
%&= \int_0^1 \mathrm{d}t \sum_{k\ge 0} \frac{k + 1}{k +2} \partial_{\mu_1} \ldots \partial_{\mu_k} \left( \Phi \frac{\delta}{\Phi_{\mu_1\ldots\mu_k \nu}} \frac{\partial \boldsymbol{\mathbf{S}}_{\bar{F}}[t\Phi]}{\partial \mathrm{d}x^\nu} \right)
%\end{split}
%\end{equation} The integration with respect to $t$ can be readily done because $\mathbf{S}_{\bar{F}}$ is linear and homogeneous in the fields (the field dependence
%comes only from the Euler-Lagrange derivatives of $L$). We obtain
%\begin{equation}
%\mathbf{K}_{\bar{F}}[\Phi] = \sum_{k\ge 0} \frac{k + 1}{k +2} \partial_{\mu_1} \ldots \partial_{\mu_k} \left( \Phi \frac{\delta}{\Phi_{\mu_1\ldots\mu_k \nu}} \frac{\partial \boldsymbol{\mathbf{S}}_{\bar{F}}[t\Phi]}{\partial \mathrm{d}x^\nu} \right)
%\end{equation} 

\paragraph{Example} Let us return to our example of general relativity. The exact reducibility parameters of the theory are the diffeomorphism generators $\bar{\xi}$, which satisfy
\begin{equation}
\delta_{\bar{\xi}} g_{\mu\nu} = \mathcal{L}_{\bar{\xi}} g_{\mu\nu} = 0 ,
\end{equation} i.e. they are the Killing vectors of $g_{\mu\nu}$. Note that for a generic metric, this equation does not admit any solution. Hence, the previous construction is irrelevant for this general case. Now, consider linearized general relativity around a background $g_{\mu\nu} = \bar{g}_{\mu\nu} + h_{\mu\nu}$. We can show that
\begin{equation}
\delta_{\bar{\xi}} h_{\mu\nu} = \mathcal{L}_{\bar{\xi}}\bar{g}_{\mu\nu} = 0 ,
\label{killing eq background}
\end{equation} i.e. the exact reducibility parameters of the linearized theory are the Killing vectors of the background $\bar{g}_{\mu\nu}$. If $\bar{g}_{\mu\nu}$ is taken to be the Minkowski metric, then the solutions of \eqref{killing eq background} are the generators of the Poincar\'e transformations. The $(n-2)$-form \eqref{DEF OF K} can be constructed explicitly and integrated on a $(n-2)$-sphere at infinity. This gives the ADM charges of linearized gravity \cite{Barnich:2001jy}.

%\paragraph{Remark} The above example suggests to extend the previous results to the \textit{linearized gauge theories}, i.e. 
%
%Asympotitcs in traeted as limit of linearized theory...
%
%
%A further extension treated in details in \cite{Barnich:2001jy} enables to study a subclass of asymptotics. Consider a field of the form $\Phi = \bar{\Phi} + \varphi$ where $\varphi$ can be arbitrary large in the bulk spacetime, but such that $\varphi / \bar{\Phi} \underbrace{\longrightarrow}_{r\to \infty} 0$ while $r \longrightarrow \infty$ ($r$ is a positive parameter such that $r=\infty$ on the spacetime boundary). The Dirichlet boundary conditions \eqref{as ads} with \eqref{BC Dirichlet} for asymptotically AdS spacetimes and the definition \eqref{asymp flat 1} with \eqref{asymp flat 3} for asymptotically flat spactimes enter in this subclass of asymptotics. \textbf{NOOOOOOOOOON?????  HOW TO GENERATE BMS SYMMETRIES SINCE SYMMETRIES OF THE BACKGROUND????}. In this framework, we can define a notion of \textit{asymptotic reducibility parameters} as field-independent gauge parameters $\bar{F}$ such $R[\bar{F}] \longrightarrow 0$ with a certain fall-off while $r \longrightarrow 0$. ........

\subsection{Asymptotic symmetries and surface charges}

We now come to the case of main interest, where we are dealing with asymptotic symmetries in the sense of the definition in subsection \ref{Asymptotic symmetry algebra}. The prescription to construct the $(n-2)$-form $\mathbf{k}_F [\Phi, \delta\Phi]$ associated with generators of asymptotic symmetries $F$ is essentially the same as the one introduced above for exact reducibility parameters. However, this $(n-2)$-form will not be conserved on-shell. Indeed, for a generic asymptotic symmetry, \eqref{weakly vanishing is conserved} does not hold; therefore, the weak equality in \eqref{weakly vanishing is conserved conseq} is not valid anymore. Nonetheless, as we will see below, we still have a control on the \textit{breaking} in the conservation law.  

\paragraph{Definition \normalfont[Barnich-Brandt $(n-2)$-form for asymptotic symmetries]} The $(n-2)$-form $\mathbf{k}_F$ associated with asymptotic symmetries generated by $F$ is defined as
\begin{equation}
\mathbf{k}_F [\Phi ; \delta \Phi] = -I^{n-1}_{\delta \Phi} \mathbf{S}_F ,
\label{expression k}
\end{equation} where $I^{n-1}_{\delta \Phi}$ is the homotopy operator \eqref{homotopy operator} and $\mathbf{S}_F$ is the weakly vanishing Noether current defined in the last line of \eqref{Noether identity proof}. For a \textit{first order gauge theory}, namely a gauge theory involving only first order derivatives of the gauge parameters $F=(f^\alpha)$ in the gauge transformations as in \eqref{First order} and first order equations of motion for the fields $\Phi= (\phi^i)$, the $(n-2)$-form \eqref{expression k} becomes
\begin{equation}
\mathbf{k}_F [\Phi ; \delta \Phi] = -\frac{1}{2} \delta \Phi \frac{\partial}{\partial (\partial_\mu \Phi)} \left( \frac{\partial}{\partial \mathrm{d}x^\mu} \mathbf{S}_F  \right) ,
\end{equation} where 
\begin{equation}
\mathbf{S}_F = R^\mu_\alpha f^\alpha \frac{\delta L}{\delta \Phi} (\mathrm{d}^{n-1}x)_\mu .
\end{equation} The simplicity of these expressions motivates the study of first order formulations of gauge theories in this context \cite{Barnich:2016rwk , Barnich:2019vzx ,  Frodden:2019ylc}.

\paragraph{Example} Let us consider the theory of general relativity. Applying the homotopy operator on the weakly vanishing Noether current $\mathbf{S}_\xi$ obtained in equation \eqref{weakly vanishin noether current gravity}, we deduce the explicit expression
\begin{equation}
\begin{split}
\mathbf{k}_\xi [g;h ] = \frac{\sqrt{-g} }{8\pi G} (\mathrm{d}^{n-2}x)_{\mu\nu} [&\xi^\nu \nabla^\mu h + \xi^\mu \nabla_\sigma h^{\sigma \nu} + \xi_\sigma \nabla^\nu h^{\sigma \mu}  \\
&+ \frac{1}{2} ( h \nabla^\nu \xi^\mu + h^{\mu \sigma} \nabla_\sigma \xi^\nu + h^{\nu\sigma} \nabla^\mu \xi_\sigma) ] ,
\end{split}
\label{Barnich-Brandt}
\end{equation} where $h_{\mu\nu} = \delta g_{\mu \nu}$. Indices are lowered and raised by $g_{\mu\nu}$ and its inverse, and $h = {h^{\mu}}_\mu$ \cite{Barnich:2001jy}. Notice that this expression has also been derived in the first order Cartan formulation of general relativity in \cite{Barnich:2016rwk} (see also \cite{Barnich:2019vzx}).

\paragraph{Theorem \normalfont[Conservation law]} Define the \textit{invariant presymplectic current as}
\begin{equation}
\mathbf{W}[\Phi;\delta \Phi , \delta \Phi ] = \frac{1}{2} I^{n}_{\delta \Phi} \left( \delta \Phi \frac{\delta \mathbf{L}}{\delta \Phi} \right) .
\label{def of invariant presympletcic potential}
\end{equation} We have the following \textit{conservation law}
\begin{equation}
\mathrm{d} \mathbf{k}_F[\Phi;\delta \Phi] \approx \mathbf{W}[\Phi; R[F] , \delta \Phi] ,
\label{breaking in the conservation}
\end{equation} where, in the equality $\approx$, it is implied that $\Phi$ is a solution of the Euler-Lagrange equations and $\delta \Phi$ is a solution of the linearized Euler-Lagrange equations. Furthermore, we use the notation $\mathbf{W} [\Phi; R[F] ,  \delta \Phi]  = i_{R[F]} \mathbf{W}[\Phi; \delta\Phi, \delta \Phi]$.

The proof of this proposition involves the properties of the operators introduced in appendix \ref{Useful results}. We have
\begin{equation}
\begin{split}
\mathrm{d} \mathbf{k}_F [\Phi;\delta\Phi] &= -\mathrm{d} I^{n-1}_{\delta \Phi} \mathbf{S}_F \\
&= \delta \mathbf{S}_F -  I^n_{\delta\Phi}\mathrm{d} \mathbf{S}_F \\
&\approx  -I^n_{\delta\Phi} \mathrm{d} \mathbf{S}_F \\
&\approx -I^n_{\delta\Phi} \left( R[F] \frac{\delta \mathbf{L}}{\delta \Phi} \right) \\
&\approx  \frac{1}{2} i_{R[F]} I^{n}_{\delta \Phi} \left( \delta \Phi \frac{\delta \mathbf{L}}{\delta \Phi} \right) \\
&\approx  i_{R[F]} \mathbf{W}[\Phi;\delta \Phi , \delta \Phi ] \\
&\approx \mathbf{W}[\Phi; R[F] ,  \delta \Phi] .
\end{split}
\end{equation} In the second equality, we used \eqref{commutation homotopy op}. In the third equality, we used the fact that $\delta \mathbf{S}_F \approx 0$, since $\delta \Phi$ is a solution of the linearized Euler-Lagrange equations. In the fourth equality, we used Noether's second theorem \eqref{second Noether theorem}. In the fifth equality, we used 
\begin{equation}
i_{R[F]} \mathbf{W}[\Phi;\delta\Phi, \delta\Phi]=  I^{n}_{R[F]} \left( \delta \Phi \frac{\delta \mathbf{L}}{\delta \Phi} \right) = - I^{n}_{\delta \Phi} \left( R[F] \frac{\delta \mathbf{L}}{\delta \Phi} \right) .
\end{equation} The proof of this statement can be found in appendix A.5 of \cite{Barnich:2007bf}. Finally, in the sixth equality, we used the definition \eqref{def of invariant presympletcic potential}.

%\paragraph{Definition \normalfont[Integrability]} The $(n-2)$-form $k_F$ is said to be \textit{integrable} if it is $\delta$-exact, namely $k_F[\Phi; \delta \Phi] = \delta K_F[\Phi]$. Notice that a field-dependent redefinition of the parameters may be necessary to render the potential integrability of the $(n-2)$-form manifest.

\paragraph{Definition \normalfont[Surface charges]} Let $\Sigma$ be a $(n-1)$-surface and $\partial\Sigma$ its $(n-2)$-dimensional boundary. We define the \textit{infinitesimal surface charge} $\ndelta H_F[\Phi]$ as
\begin{equation}
\ndelta H_F[\Phi] = \int_{\partial\Sigma} \mathbf{k}_F[\Phi;\delta\Phi] \approx \int_\Sigma \mathbf{W}[\Phi; R[F] ,  \delta \Phi] .
\label{inifnitesimal surf charge}
\end{equation} The infinitesimal surface charge $\ndelta H_F[\Phi]$ is said to be \textit{integrable} if it is $\delta$-exact, i.e. if\\$\ndelta H_F[\Phi] = \delta H_F [\Phi]$. The symbol $\ndelta$ in \eqref{inifnitesimal surf charge} emphasizes that the infinitesimal surface charge is not necessarily integrable. If it is actually integrable, then we can define the \textit{integrated surface charge} $H_F[\Phi]$ as
\begin{equation}
H_F [\Phi] = \int_\gamma \delta H_F [\Phi]  + N[\bar{\Phi}] = \int_\gamma
 \int_{\partial\Sigma} \mathbf{k}_F[\Phi;\delta\Phi]+ N[\bar{\Phi}] ,
 \label{integrated charge}
\end{equation} where $\gamma$ is a path in the solution space, going from a reference solution $\bar{\Phi}$ to the solution $\Phi$. $N[\bar{\Phi}]$ is a chosen value of the charge for this reference solution, which is not fixed by the formalism. Notice that for integrable infinitesimal charge, the integrated charge $H_F [\Phi]$ is independent from the chosen path $\gamma$ \cite{Compere:2019qed}. 

\paragraph{Theorem \normalfont[Charge representation theorem]} Assuming integrability, the integrated surface charges satisfy the algebra
\begin{equation}
\{ H_{F_1} , H_{F_2} \} \approx H_{[F_1, F_2]_A} + K_{F_1; F_2}[\bar{\Phi}] .
\label{charge algebra integrable}
\end{equation} In this expression, the integrated charges bracket is defined as
\begin{equation}
\{ H_{F_1} , H_{F_2} \} = \delta_{F_2} H_{F_1} =\int_{\partial \Sigma} \mathbf{k}_{F_1} [\Phi;\delta_{F_2} \Phi] .
\end{equation} Furthermore, the central extension $K_{F_1;F_2}[\bar{\Phi}]$, which depends only on the reference solution $\bar{\Phi}$, is antisymmetric with respect to $F_1$ and $F_2$, i.e. $K_{F_1;F_2}[\bar{\Phi}] = K_{F_2;F_1}[\bar{\Phi}]$. It satisfies the $2$-cocycle condition
\begin{equation}
K_{[F_1, F_2]_A ;F_3}[\bar{\Phi}] + K_{[F_2, F_3]_A ;F_1}[\bar{\Phi}] + K_{[F_3, F_1]_A ;F_2}[\bar{\Phi}] \approx 0 .
\end{equation} Therefore, the integrated charges form a representation of the asymptotic symmetry algebra, up to a central extension \cite{Barnich:2001jy, Barnich:2007bf}. 

For the proof of this theorem, see e.g. section 1.4 of \cite{Compere:2019qed}.

\paragraph{Remark} In the literature, there are several criteria based on properties of the surface charges, that make a choice of boundary conditions interesting. The main properties are the following:
\begin{itemize}
\item The charges are usually required to be \textit{finite}. Two types of divergences may occur: divergences in the expansion parameter defining asymptotics, say $r$, and divergences when performing the integration on the $(n-2)$-surface $\partial \Sigma$.
\item The charges have to be \textit{integrable}. As explained above, this criterion enables us to define integrated surface charges as in \eqref{integrated charge}. Integrability implies that the charges form a representation of the asymptotic symmetry algebra, up to a central extension (see \eqref{charge algebra integrable}). Furthermore, the integrated charges generate the symmetries on the solution space. 
\item The charges have to be \textit{generically non-vanishing}. Indeed, since the integrated surface charges generate the symmetry, identically vanishing charges would imply trivial action on the solution space. In particular, the asymptotic symmetries for which associated integrated charges identically vanish are considered as trivial in the strong definition of asymptotic symmetry group \eqref{ASG def 3}.
\item The charges have to be \textit{conserved} in time when the integration is performed on a spacelike $(n-2)$-dimensional surface $\partial \Sigma$ at infinity. This statement is not guaranteed a priori because of the breaking in the conservation law \eqref{breaking in the conservation}. 
\end{itemize}

However, even if these requirements seem reasonable, in practice, some of them may not be satisfied. Indeed, as we will see below, the BMS charges in four dimensions are not always finite, neither integrable, nor conserved \cite{Barnich:2011mi}. We now discuss the violation of some of the above requirements:
\begin{itemize}
\item The fact that the charges may \textit{not be finite} in terms of the expansion parameter $r$ can be expected when the asymptotic region is taken to be at infinity. Indeed, consider $r$ as a cut-off. It makes sense to integrate on a surface $\partial\Sigma$ at a constant finite value of $r$, encircling a finite volume. Then, taking the limit $r\to \infty$ leads to an infinite volume; therefore, it does not come as a surprise that quantities diverge. Furthermore, it has recently been shown in \cite{Godazgar:2018vmm} that subleading orders in $r$ in the $(n-2)$-form $\mathbf{k}_{F}[\Phi;\delta \Phi]$ contain some interesting physical information, such as the $10$ conserved Newman-Penrose charges \cite{Newman:1968uj}. Therefore, it seems reasonable to think that overleading orders in $r$ may also contain relevant information (see e.g. \cite{Compere:2017wrj , Compere:2019odm}). 
\item The \textit{non-integrability} of the charges may be circumvented by different procedures to isolate an integrable part in the expression of the charges (see e.g. \cite{Wald:1999wa} and \cite{Compere:2018ylh}). However, the final integrated surface charges obtained by these procedures do not have all the properties that integrable charges would have. In particular, the representation theorem does not generically hold. Another philosophy is to keep working with non-integrable expressions, without making any specific choice for the integrable part of the charges. In some situations, it is still possible to define a modified bracket for the charges, leading to a representation of the asymptotic symmetry algebra, up to a $2$-cocycle which may depend on fields \cite{Barnich:2011mi, Compere:2018ylh }. However, no general representation theorem exists in this context, even if some progress has been made \cite{Troessaert:2015nia}.
\item Finally, the \textit{non-conservation} of the charges contains some important information on the physics. For example, in asymptotically flat spacetimes at null infinity, the non-conservation in time of the charges associated with time translations is known as the \textit{Bondi mass loss}. This tells us that the mass decreases in time at future null infinity because of a flux of radiation through the boundary. Hence, the non-conservation of the charges contains important information on the dynamics of the system.      
\end{itemize} Even if the charges have these pathologies, they still offer important insights on the system. They could be seen as interesting combinations of the elements of the solution space that enjoy some properties in their transformation (see e.g. \cite{Barnich:2016lyg, Barnich:2015uva}). 

\paragraph{Examples} We now provide explicit examples of surface charge constructions in four-dimensional general relativity. First, consider asymptotically AdS$_4$ spacetimes with Dirichlet boundary conditions (AAdS2) (condition \eqref{as ads} together with \eqref{BC Dirichlet}), the associated solution derived in subsection \ref{Solution space} (equation \eqref{solution space dirichlet}), and the associated asymptotic Killing vectors derived in subsection \ref{Asymptotic symmetry algebra}. Inserting this solution space and these asymptotic Killing vectors into the $(n-2)$-form \eqref{Barnich-Brandt} results in an integrable expression at order $\rho^0$. Therefore, we can construct an integrated surface charge \eqref{integrated charge} where the $2$-surface $\partial \Sigma$ is taken to be the $2$-sphere at infinity, written $S_\infty$. We have the explicit expression
\begin{equation}
H_\xi [g] = \int_{S_\infty} \mathrm{d}^2\Omega ~(\xi_0^a {T_a}^t  ) ,
\end{equation} where $\mathrm{d}^2\Omega$ is the integration measure on the $2$-sphere (see e.g. \cite{Compere:2008us}). These charges are finite and generically non-vanishing. Furthermore, we can easily show that they are conserved in time, i.e.
\begin{equation}
\frac{{d}}{{d}t} H_\xi [g] \approx 0 .
\end{equation}

Now, we consider definition \eqref{asymp flat 1} with \eqref{asymp flat 3} of asymptotically flat spacetimes in four dimensions (AF3). The surface charges are obtained by inserting the corresponding solution space derived in subsection \ref{Solution space} (see equation \eqref{solution space Bondi dirich}) and the asymptotic Killing vectors discussed in subsection \ref{Asymptotic symmetry algebra} into the expression \eqref{Barnich-Brandt}, and then integrating over $S_\infty$. The result is given by
\begin{equation}
\ndelta H_\xi [g;\delta g] \approx \delta J_\xi [g] + \Theta_\xi [g; \delta g] ,
\label{BMS charges}
\end{equation} where
\begin{equation}
\begin{split}
&J_\xi [g] = \frac{1}{16\pi G} \int_{S_\infty} \mathrm{d}^2 \Omega ~\left[4 f M + Y^A (2N_A + \frac{1}{16} \partial_A (C^{CB} C_{CB}) ) \right] \\
&\Theta_\xi [g;\delta g] = \frac{1}{16 \pi G} \int_{S_\infty} \mathrm{d}^2 \Omega ~\left[\frac{f}{2} N_{AB} \delta C^{AB} \right]
\end{split}
\label{integrable and non integrable parts}
\end{equation} and where $N_{AB} = \partial_u C_{AB}$ \cite{Barnich:2011mi}. As mentioned above, the infinitesimal surface charges are not integrable. Therefore, we cannot unambiguously define an integrated surface charge as in \eqref{integrated charge} (see, however, \cite{Wald:1999wa , Compere:2018ylh}). In particular, the representation theorem \eqref{charge algebra integrable} does not hold. Nevertheless, we can define the following modified bracket \cite{Barnich:2011mi}:
\begin{equation}
\{ J_{\xi_1} , J_{\xi_2} \}^* = \delta_{\xi_2} J_{\xi_1}[g] + \Theta_{\xi_2} [g;\delta_{\xi_1}g ] .
\end{equation} We can show that
\begin{equation}
\{ J_{\xi_1} , J_{\xi_2} \}^* \approx J_{[\xi_1, \xi_2]_A} [g] +  K_{\xi_1;\xi_2}[g] ,
\label{algebra modified}
\end{equation} where $ K_{\xi_1;\xi_2}[g]$ is a \textit{field-dependent} $2$-cocycle given explicitly by\footnote{Notice that this $2$-cocycle is zero for globally well-defined conformal transformations on the $2$-sphere. It becomes non-trivial when considering the extended BMS$_4$ group with Vir $\times$ Vir superrotations.}
\begin{equation}
K_{\xi_1;\xi_2}[g] = \frac{1}{32 \pi G} \int_{S_\infty} \mathrm{d}^2 \Omega~ [C^{BC} (f_1 D_B D_C D_A Y^A_2 - f_2 D_B D_C D_A Y_1^A)] .
\end{equation} It satisfies the generalized $2$-cocycle condition 
\begin{equation}
K_{[\xi_1, \xi_2]_A, \xi_3} + \delta_{\xi_3} K_{\xi_1, \xi_2} + \text{cyclic (1,2,3)} \approx 0 .
\label{generalized 2 cocycle condition}
\end{equation} For the algebra \eqref{algebra modified} to make sense, its form should not depend on the particular choice of integrable part $J_\xi[g]$ in \eqref{integrable and non integrable parts}. In particular, defining $J' = J - N$ and $\Theta' = \Theta + \delta N$ for some $N = N_\xi[g]$, we obtain
\begin{equation}
\{ J'_{\xi_1} , J'_{\xi_2} \}^* = J'_{[\xi_1, \xi_2]_A} [g] +  K'_{\xi_1;\xi_2}[g] ,
\end{equation} where $\{ J'_{\xi_1} , J'_{\xi_2} \}^*= \delta_{\xi_2} J'_{\xi_1}[g] + \Theta'_{\xi_2}[g;\delta_{\xi_1} g]$ and 
\begin{equation}
K'_{\xi_1;\xi_2} = K_{\xi_1, \xi_2} - \delta_{\xi_2} N_{\xi_1} + \delta_{\xi_1} N_{\xi_2} + N_{[\xi_1, \xi_2]_A} .
\end{equation} Notice that $- \delta_{\xi_2} N_{\xi_1} + \delta_{\xi_1} N_{\xi_2} + N_{[\xi_1, \xi_2]_A}$ automatically satisfies the generalized $2$-cocycle condition \eqref{generalized 2 cocycle condition} \cite{Barnich:2011mi}. Another property of the surface charges \eqref{BMS charges} and \eqref{integrable and non integrable parts} is that they are not conserved. Indeed, 
\begin{equation}
\frac{d}{du} \ndelta H_{\xi}[g] = \int_{S_\infty} \mathbf{W}[g;\delta_\xi g , \delta g] ,
\label{bondi mass loss}
\end{equation} where $\mathbf{W}[g;\delta g , \delta g]$ was computed\footnote{More precisely, in \cite{Compere:2018ylh}, we computed the presymplectic potential $\boldsymbol{\omega}[g;\delta g , \delta g ]$ introduced below. However, as we will see, this is equal to the invariant presymplectic current in the Bondi gauge.} in \cite{Compere:2018ylh}. We have
\begin{equation}
\int_{S_\infty}  \mathbf{W}[g;\delta g , \delta g] = -\frac{1}{32\pi G}\int_{S_\infty} \mathrm{d}^2 \Omega~ [\delta N^{AB} \wedge \delta C_{AB} ] .
\label{non conservation BMS}
\end{equation} Notice that taking $f = 1$ and $Y^A =0$ in \eqref{bondi mass loss}, we recover the famous \textit{Bondi mass loss formula} \cite{Bondi:1962px , Sachs:1962wk , Sachs:1962zza}. This formula indicates that the mass is decreasing in time because of the leak of radiation through $\mathscr{I}^+$. This was a striking argument for the existence of gravitational waves at the non-linear level of the theory. Finally, despite the BMS charges \eqref{BMS charges} and \eqref{integrable and non integrable parts} not being divergent in $r$, we can show that some of the supertranslation charges diverge for the Kerr solution \cite{Barnich:2011mi}.

\paragraph{Remark} A non-trivial relation seems to exist between conservation and integrability of the surface charges. For example, in the case of Dirichlet boundary conditions in asymptotically AdS$_4$ spacetimes (AAdS2) considered above, we see that the surface charges are both integrable and conserved. Reciprocally, there is a relation between non-conservation and non-integrability of the surface charges. For example, in the asymptotically flat case (AF3), we see that the source of non-integrability is contained in the asymptotic shear $C_{AB}$ and the news function $N_{AB} = \partial_u C_{AB}$. These are precisely the functions involved in the right-hand side of \eqref{non conservation BMS}. We can consider many other examples where this phenomenon appears. Therefore, non-integrability is related to non-conservation of the charges. We will see below that for diffeomorphism-invariant theories, the relation between non-conservation and integrability is transparent in the covariant phase space formalism.

\subsection{Relation between Barnich-Brandt and Iyer-Wald procedures}

In this subsection, we briefly discuss the covariant phase space formalism leading to the Iyer-Wald prescription for surface charges \cite{Wald:1999wa, Iyer:1994ys , Wald:1993nt , Harlow:2019yfa}. Notice that this method is valid only for diffeomorphism-invariant theories (including general relativity), and not for any gauge theories. In practice, this means that the parameters of the asymptotic symmetries are diffeomorphisms generators, i.e. $F \equiv \xi$ and $\delta_F \Phi \equiv \mathcal{L}_\xi \Phi$. Finally, we relate this prescription to the Barnich-Brandt prescription presented in detail in the previous section.

\paragraph{Definition \normalfont[Presymplectic form]} Consider a diffeomorphism-invariant theory with Lagrangian $\mathbf{L} = L \mathrm{d}^n x$. Let us perform an arbitrary variation of the Lagrangian. Using a similar procedure as in \eqref{stage 2}, we obtain
\begin{equation}
\begin{split}
\delta L &= \delta \Phi \frac{\partial L}{\partial \Phi} + \delta \partial_\mu \Phi   \frac{\partial L}{\partial (\partial_\mu \Phi)} + \ldots \\
&= \delta \Phi \frac{\delta L}{\delta \Phi} + \partial_\mu \left( \delta \Phi \frac{\partial L}{\partial (\partial_\mu \Phi ) } + \ldots \right) \\
&= \delta \Phi \frac{\delta L}{\delta \Phi} + \partial_\mu  \theta^\mu [\Phi; \delta \Phi] ,
\end{split}
\label{eq:presymp construct}
\end{equation} where 
\begin{equation}
\boldsymbol{\theta} [ \Phi; \delta \Phi ] = \theta^\mu [ \Phi; \delta \Phi ] (\mathrm{d}^{n-1}x)_\mu = \left( \delta \Phi \frac{\partial L}{\partial (\partial_\mu \Phi ) } + \ldots \right) (\mathrm{d}^{n-1}x)_\mu = I^n_{\delta \Phi} \mathbf{L}
\label{presymplectic potential}
\end{equation} is the \textit{presymplectic potential}. Taking into account that $\delta$ is Grassmann odd, the equation \eqref{eq:presymp construct} can be rewritten as
\begin{equation}
\delta \mathbf{L} = \delta \Phi \frac{\delta \mathbf{L}}{\delta \Phi} - \mathrm{d} \boldsymbol{\theta} [\Phi; \delta \Phi] .
\label{var lagrangien presymp}
\end{equation} Now, the \textit{presymplectic form} $\boldsymbol{\omega}$ is defined as
\begin{equation}
\boldsymbol{\omega}[\Phi;\delta \Phi, \delta \Phi] = \delta \boldsymbol{\theta} [\Phi, \delta \Phi ] .
\label{presymplectic form}
\end{equation}

\paragraph{Definition \normalfont[Iyer-Wald $(n-2)$-form for asymptotic symmetries]} The \textit{Iyer-Wald $(n-2)$-form} $\mathbf{k}^{IW}_\xi$ associated with asymptotic symmetries generated by $\xi$ is defined as
\begin{equation}
\mathbf{k}_\xi^{IW} [\Phi; \delta \Phi] = -\delta \mathbf{Q}_\xi [\Phi] + \iota_{\xi} \boldsymbol{\theta}[\Phi ; \delta \Phi ] ,
\label{Iyer-Wald}
\end{equation} up to an exact $(n-2)$-form\footnote{In the definition \eqref{Iyer-Wald}, we assumed that the variational operator $\delta$ in front of the Noether-Wald charge does not see the possible field-dependence of the asymptotic Killing vectors $\xi^\mu$. Strictly speaking, one should write $\mathbf{k}_\xi^{IW} [\Phi; \delta \Phi] = -\delta \mathbf{Q}_\xi [\Phi] +  \mathbf{Q}_{\delta \xi} [\Phi] - \iota_{\xi} \boldsymbol{\theta}[\Phi ; \delta \Phi ]$.}. In this expression, $\mathbf{Q}_\xi [\Phi]  = - I_\xi^{n-1} \boldsymbol{\theta} [\Phi; \mathcal{L}_\xi \Phi]$ is called the \textit{Noether-Wald} surface charge. 

\paragraph{Example} For general relativity theory, the presymplectic potential \eqref{presymplectic potential} is given by 
\begin{equation}
\boldsymbol{\theta}[g;h] =  \frac{\sqrt{-g}}{16\pi G}  (\nabla_\nu h^{\mu\nu} - \nabla^\mu h )(\mathrm{d}^{n-1}x)_\mu ,
\end{equation} where $h_{\mu\nu} = \delta g_{\mu \nu}$. Indices are lowered and raised by $g_{\mu\nu}$ and its inverse, and $h = {h^{\mu}}_\mu$. From this expression, the Noether-Wald charge can be computed; we obtain
\begin{equation}
\mathbf{Q}_\xi [g]  = - I_\xi^{n-1} \boldsymbol{\theta} [g; \mathcal{L}_\xi g] = \frac{\sqrt{-g}}{8\pi G}  \nabla^\mu \xi^\nu (\mathrm{d}^{n-2} x)_{\mu\nu}
\end{equation} and we recognize the \textit{Komar charge}. Finally, inserting these expression into \eqref{Iyer-Wald} yields
\begin{equation}
\mathbf{k}_\xi^{IW} [g;h] = \frac{\sqrt{-g}}{8 \pi G}  \left(   \xi^\mu \nabla_\sigma h^{\nu\sigma} - \xi^\mu \nabla^\nu h + \xi_\sigma \nabla^\nu h^{\mu\sigma} + \frac{1}{2} h \nabla^\nu \xi^\mu - h^{\rho\nu} \nabla_\rho \xi^\mu \right) (\mathrm{d}^{n-2} x)_{\mu\nu} .
\label{IW for GR}
\end{equation}

\paragraph{Theorem \normalfont[Conservation law]} We have the following \textit{conservation law}:
\begin{equation}
\mathrm{d} \mathbf{k}^{IW}_\xi [\Phi;\delta \Phi] \approx \boldsymbol{\omega} [\Phi;  \mathcal{L}_\xi \Phi, \delta \Phi] ,
\label{breaking in the conservation 2}
\end{equation} where, in the equality $\approx$, it is implied that $\Phi$ is a solution of the Euler-Lagrange equations and $\delta \Phi$ is a solution of the linearized Euler-Lagrange equations. Furthermore, $\boldsymbol{\omega} [\Phi; \mathcal{L}_\xi \Phi,  \delta \Phi ]  =  i_{\mathcal{L}_\xi \Phi} \boldsymbol{\omega}[\Phi; \delta\Phi, \delta \Phi]=- \boldsymbol{\omega}[\Phi; \delta \Phi ,  \mathcal{L}_\xi \Phi]$. 

This can be proved using Noether's second theorem \eqref{second Noether theorem} (see e.g. \cite{Compere:2019qed} for a detailed proof). 

\paragraph{Remark} In the covariant phase space formalism, the relation between non-integrability and non-conservation mentioned in the previous subsection is clear. Indeed, 
\begin{equation}
\begin{split}
\delta \ndelta H_\xi [\Phi] &=  \int_{\partial \Sigma} \delta \textbf{k}^{IW}_\xi [\Phi, \delta \Phi] \\
&= \int_{\partial \Sigma} \delta \iota_\xi \boldsymbol{\theta} [g, \delta g] \\
&= \int_{\partial \Sigma} -\iota_\xi  \delta \boldsymbol{\theta} [g, \delta g] \\
&= \int_{\partial \Sigma}- \iota_\xi \boldsymbol{\omega} [g;\delta g, \delta g ] ,
\end{split}
\end{equation} where we used \eqref{Iyer-Wald} and \eqref{presymplectic form} in the second and the fourth equality, respectively. The surface charge $\ndelta H_\xi [\Phi]$ is integrable only if $\delta \ndelta H_\xi [\Phi] =0 $, if and only if
\begin{equation}
\int_{\partial \Sigma} \iota_\xi \boldsymbol{\omega} [g;\delta g, \delta g ] = 0
\end{equation} Therefore, from 
\begin{equation}
\mathrm{d} \ndelta H_\xi [\Phi] =  \int_{\partial \Sigma} \mathrm{d} \mathbf{k}^{IW}_\xi [g,\delta g] \approx \int_{\partial \Sigma} \boldsymbol{\omega} [\Phi;  \mathcal{L}_\xi \Phi, \delta \Phi] ,
\end{equation} the non-conservation is controlled by $\boldsymbol{\omega}[g,\delta g , \delta g]$ and is an obstruction for the integrability.

\paragraph{Remark} As in the Barnich-Brandt procedure, the Iyer-Wald $(n-2)$-form \eqref{Iyer-Wald} is defined up to an exact $(n-2)$-form. However, there is another source of ambiguity here coming from the definition of the presymplectic potential \eqref{presymplectic potential}. In fact, we have the freedom to shift $\boldsymbol{\theta}$ by an exact $(n-1)$-form as
\begin{equation}
\boldsymbol{\theta}[\Phi; \delta \Phi]\to \boldsymbol{\theta}[\Phi; \delta \Phi] - \mathrm{d} \mathbf{Y}[\Phi; \delta \Phi] ,
\end{equation} where $\mathbf{Y}[\Phi; \delta \Phi]$ is a $(n-2)$-form. This implies that the presymplectic form \eqref{presymplectic form} is modified as
\begin{equation}
\boldsymbol{\omega}[\Phi; \delta \Phi, \delta \Phi] \to \boldsymbol{\omega} [\Phi; \delta \Phi, \delta \Phi] + \mathrm{d} \delta \mathbf{Y}[\Phi; \delta \Phi] ,
\end{equation} where we used the fact that both $\mathrm{d}$ and $\delta$ are Grassmann odd. The Noether-Wald charge becomes
\begin{equation}
\mathbf{Q}_\xi [\Phi] \to \mathbf{Q}_\xi[\Phi] +  \mathbf{Y}[\Phi; \mathcal{L}_\xi \Phi] ,
\end{equation} up to an exact $(n-2)$-form which can be reabsorbed in the $(n-2)$-form ambiguity for $\mathbf{k}^{IW}_\xi$ discussed above. Therefore, this ambiguity modifies $\mathbf{k}^{IW}_{F}$ given in \eqref{Iyer-Wald} by
\begin{equation}
\mathbf{k}^{IW}_\xi [\Phi; \delta \Phi] \to \mathbf{k}^{IW}_\xi [\Phi; \delta \Phi] - \delta \mathbf{Y}[\Phi; \mathcal{L}_\xi \Phi] - \iota_\xi \mathrm{d} \mathbf{Y}[\Phi; \delta \Phi] .
\label{modif on n-2}
\end{equation} 

%\textbf{changer en dessous, autre ambig!!!!}
%
%Notice that a particular way to bring this ambiguity is to modify the Lagrangian by a boundary term $\mathbf{L} \to \mathbf{L} + \mathbf{d} \mathbf{M}$ where $\mathbf{M}$ is a $(n-1)$ form. In this case, we have $\mathbf{Y}[\Phi;\delta \Phi] = \delta \mathbf{M}[\Phi]$ and, therefore, this has no influence on the presymplectic form $\boldsymbol{\omega}[\Phi; \delta \Phi, \delta \Phi]$, neither on the $(n-2)$-form. Indeed, in this case, the modification \eqref{modif on n-2} takes the form 
%\begin{equation}
%\begin{split}
%- \delta \mathbf{Y}[\Phi; \delta_\xi \Phi] + \iota_\xi d \mathbf{Y}[\Phi; \delta \Phi] &= -\delta \delta_\xi \textbf{M} + \iota_\xi d \delta \textbf{M} \\
%&= -\delta (\mathcal{L}_\xi \textbf{M} ) + \mathcal{L}_\xi \delta \textbf{M} + d \iota_\xi \delta \textbf{M} \\
%&=  d \iota_\xi \delta \textbf{M} 
%\end{split}
%\end{equation} In the second equality, we use the Cartan magic formula $\mathcal{L}_\xi = d \iota_\xi + \iota_\xi d$. In the last equality, we used the second relation of \eqref{commutation relation delta Q}. Therefore, the $(n-2)$-form is modified only up to an exact $(n-2)$-form.

\paragraph{Definition} Let us introduce an important $(n-2)$-form which is involved in the relation between the Barnich-Brandt and Iyer-Wald prescriptions discussed in the remark below. We define
\begin{equation}
\mathbf{E}[\Phi; \delta \Phi , \delta \Phi] = - \frac{1}{2} I^{n-1}_{\delta \Phi} \boldsymbol{\theta} = - \frac{1}{2} I^{n-1}_{\delta \Phi} I^n_{\delta \Phi} \mathbf{L} .
\label{definition}
\end{equation} 

\paragraph{Remark} We now relate the Barnich-Brandt and the Iyer-Wald prescriptions to construct the $(n-2)$-form. Let us start from the expression \eqref{var lagrangien presymp} of the variation of the Lagrangian. We apply the homotopy operator on each side of the equality. We have
\begin{equation}
\begin{split}
I_{\delta \Phi}^n \delta \mathbf{L} &= I^n_{\delta \Phi} \left( \delta \Phi \frac{\delta \mathbf{L}}{\delta \Phi} \right) - I^n_{\delta \Phi} \mathrm{d} \boldsymbol{\theta} \\
&= I^n_{\delta \Phi} \left( \delta \Phi \frac{\delta \mathbf{L}}{\delta \Phi} \right) - \delta \boldsymbol{\theta} - \mathrm{d} I^{n-1}_{\delta \Phi} \boldsymbol{\theta} .
\end{split}
\end{equation} Therefore,
\begin{equation}
I_{\delta \Phi}^n \delta \mathbf{L} + \delta \boldsymbol{\theta}= I^n_{\delta \Phi} \left( \delta \Phi \frac{\delta \mathbf{L}}{\delta \Phi} \right) - \mathrm{d} I^{n-1}_{\delta \Phi} \boldsymbol{\theta} .
\end{equation} 
Since $[\delta , I_{\delta \Phi}^n] = 0$ because $\delta^2 = 0$, the left-hand side of the last equality can be rewritten as $\delta I_{\delta \Phi}^n  \mathbf{L} + \delta \boldsymbol{\theta} = 2 \delta \boldsymbol{\theta} = 2 \boldsymbol{\omega}$ where we used \eqref{presymplectic potential}. Now, using \eqref{def of invariant presympletcic potential} and \eqref{definition}, we obtain the relation between the presymplectic form $\boldsymbol{\omega}$ and the invariant presymplectic current $\mathbf{W}$ as
\begin{equation}
\boldsymbol{\omega}[\Phi;\delta \Phi, \delta \Phi] = \mathbf{W}[\Phi;\delta \Phi, \delta \Phi] + \mathrm{d} \mathbf{E}[\Phi;\delta \Phi, \delta \Phi] .
\end{equation} Contracting this relation with $i_{\mathcal{L}_\xi \Phi}$ results in
\begin{equation}
\boldsymbol{\omega}[\Phi;\mathcal{L}_\xi \Phi, \delta \Phi] = \mathbf{W}[\Phi;\mathcal{L}_\xi \Phi, \delta \Phi] + \mathrm{d} \mathbf{E}[\Phi;\delta \Phi,\mathcal{L}_\xi \Phi] .
\end{equation} Finally, using the on-shell conservation laws \eqref{breaking in the conservation} and \eqref{breaking in the conservation 2}, we obtain
\begin{equation}
\mathbf{k}^{IW}_\xi [\Phi;\delta \Phi] \approx \mathbf{k}_\xi [\Phi;\delta \Phi] + \mathbf{E}[\Phi;\delta \Phi, \mathcal{L}_\xi \Phi] , 
\label{ling IW BB}
\end{equation} up to an exact $(n-2)$-form. Therefore, the Barnich-Brandt $(n-2)$-form $\mathbf{k}_\xi [\Phi;\delta \Phi]$ differs from the Iyer-Wald $(n-2)$-form $\mathbf{k}_\xi^{IW} [\Phi;\delta \Phi]$ by the term $\mathbf{E}[\Phi;\delta \Phi, \mathcal{L}_\xi \Phi]$. 

\paragraph{Examples} We illustrate these concepts with the case of general relativity. The $(n-2)$-form $\mathbf{E}[\Phi;\delta \Phi, \delta \Phi]$ can be computed using \eqref{definition}. We obtain 
\begin{equation}
\mathbf{E}[g;\delta g, \delta g] = \frac{\sqrt{-g}}{32\pi G} {(\delta g)^\mu}_\sigma \wedge (\delta g)^{\sigma \nu} (\mathrm{d}^{n-2} x)_{\mu\nu} .
\label{def en gr}
\end{equation} When contracted with $i_{\mathcal{L}_\xi g}$, this leads to
\begin{equation}
\mathbf{E}[g; \delta g , \mathcal{L}_\xi g] =- \frac{\sqrt{-g}}{16\pi G} (\nabla^\mu \xi_\sigma + \nabla_\sigma \xi^\mu) (\delta g)^{\sigma\nu} (\mathrm{d}^{n-2}x)_{\mu\nu} ,
\label{difference BB IW}
\end{equation} up to an exact $(n-2)$-form. This expression can also be obtained from \eqref{ling IW BB} by comparing the explicit expressions \eqref{Barnich-Brandt} and \eqref{IW for GR}. Notice that the difference between the Barnich-Brandt and the Iyer-Wald definitions \eqref{difference BB IW} vanishes for a Killing vectors $\xi^\mu$. Furthermore, a simple computation shows that the $(n-2)$-form \eqref{def en gr} vanishes in both the Fefferman-Graham gauge \eqref{FG gauge} and the Bondi gauge \eqref{Bondi gauge}. Therefore, the Barnich-Brandt and the Iyer-Wald prescriptions lead to the same surface charges in these gauges. For an example where the two prescriptions do not coincide, see for instance, \cite{Azeyanagi:2009wf}.

\section{Applications}
\label{Applications}

Asymptotic symmetries have a wide range of applications in theoretical physics. We briefly mention two of them and explain why the formalism presented above is relevant in these contexts. 

\subsection{Holography}

The \textit{holographic principle} states that quantum gravity can be described in terms of lower-dimensional dual quantum field theories \cite{tHooft:1993dmi , Susskind:1994vu}. A concrete realization of the holographic principle asserts that the type IIB string theory living in the bulk spacetime AdS$_5$ $\times$ $S^5$ is dual to the $\mathcal{N} = 4$ supersymmetric Yang-Mills theory living on the four-dimensional spacetime boundary \cite{Maldacena:1997re}. The gravitational theory is effectively living in the five-dimensional spacetime AdS$_5$, the five dimensions of the factor $S^5$ being compactified. A first extension of this original holographic duality is the \textit{AdS/CFT correspondence} which tells us that the gravitational theory living in the $(d+1)$-dimensional asymptotically AdS spacetime (AAdS2) is dual to a CFT living on the $d$-dimensional boundary. Other holographic dualities with different types of asymptotics have also been studied. A holographic dictionary enables one to interpret properties of the bulk theory in terms of the dual boundary theory. For example, the dictionary imposes the following relation between the symmetries of the two theories:
\begin{equation}
\left[
\begin{array}{c}
\text{Gauge symmetries in the bulk theory} \\
\Longleftrightarrow \\
\text{Global symmetries in the boundary theory}.
\end{array}
\right]
\label{relation symmetries}
\end{equation} More specifically for us, consider a given bulk solution space with asymptotic symmetries. The correspondence tells us that a set of quantum field theories exist that are associated with the bulk solutions, such that in the UV regime, the \textit{global symmetries} of these theories are exactly the \textit{asymptotic symmetries} of the bulk solution space. Even if the AdS/CFT correspondence has not been proven yet, it has been verified in a number of situations and extended in various directions. 

We now mention a famous hint in favor of this correspondence using the relation \eqref{relation symmetries}. Brown and Henneaux showed that the asymptotic symmetry group for asymptotically AdS$_3$ spacetime with Dirichlet boundary conditions is given by the infinite-dimensional group of conformal transformations in two dimensions. Furthermore, they revealed that the associated surface charges are finite, are integrable, and exhibit a non-trivial central extension in their algebra. This \textit{Brown-Henneaux central charge} is given by 
\begin{equation}
c = \frac{3 \ell}{2G} ,
\label{Brown Henneaux}
\end{equation}  where $\ell$ is the AdS$_3$ radius. The AdS/CFT correspondence tells us that there is a set of two-dimensional dual conformal field theories. The remarkable fact is that, when inserting the central charge \eqref{Brown Henneaux} into the Cardy entropy formula valid for 2$d$ CFT \cite{Strominger:1997eq}, this reproduces exactly the entropy of three-dimensional BTZ black hole solutions \cite{Banados:1992wn, Banados:1992gq}. 

The holographic principle is believed to hold in all types of asymptotics. In particular, in asymptotically flat spacetimes, from the correspondence \eqref{relation symmetries}, the dual theory would have BMS as the global symmetry. Important steps have been taken in this direction in three and four dimensions (see e.g. \cite{Dappiaggi:2004kv, Bagchi:2010eg, Barnich:2012xq , Riegler:2016hah , Fareghbal:2014qga , Bagchi:2014iea , Basu:2017aqn , Fareghbal:2013ifa} and references therein). Furthermore, in four-dimensional asymptotically flat spacetimes, traces of two-dimensional CFT seem to appear, enabling the use of well-known techniques of the AdS/CFT correspondence \cite{Barnich:2010eb , deBoer:2003vf , Ball:2019atb , Pasterski:2016qvg , Kapec:2016jld , Cheung:2016iub, Mishra:2017zan , Donnay:2018neh , Puhm:2019zbl}. Notice that global BMS symmetry can be seen as a conformal Carroll symmetry \cite{Duval:2014uva , Duval:2014lpa , Figueroa-OFarrill:2019sex}, which is especially relevant in the context of the fluid/gravity correspondence \cite{Hubeny:2011hd , Haack:2008cp, Ciambelli:2019lap, Campoleoni:2018ltl, Ciambelli:2017wou, Caldarelli:2012cm}.

\subsection{Infrared physics}

A connection has recently been established between various areas of gauge theories that are \textit{a priori} unrelated, namely \textit{asymptotic symmetries}, \textit{soft theorems} and \textit{memory effects} (see \cite{Strominger:2017zoo} for a review). These three fields of research are often referred to as the three corners of the infrared triangle of gauge theories (see figure \ref{Fig:IRSector}).

\tikzset{every node/.style={fill=white, draw=black, text width=2.5cm, align=center, inner sep=10pt}}
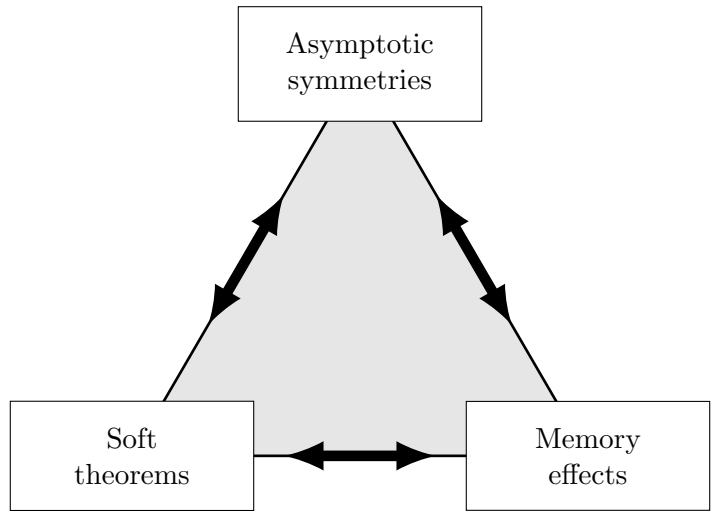
\begin{figure}[h!]
\centering
\begin{tikzpicture}
\draw[white] (-5,-3) -- (-5,4) -- (5,4) -- (5,-3) -- cycle;
\def\a{5.196};
\coordinate (A) at (-3,-2);
\coordinate (B) at (3,-2);
\coordinate (C) at (0,\a-2);
\filldraw[fill=black!10, draw=black, line width=1pt] (A) -- (B) -- (C) -- cycle;
\node[] at (A){Soft \\ theorems};
\node[] at (B){Memory effects};
\node[] at (C){Asymptotic symmetries};
\draw[line width=4pt,latex-latex] (-1,-2) -- (1,-2);
\draw[line width=4pt,latex-latex] (-2,-2+\a/3) -- (-1,-2+2*\a/3);
\draw[line width=4pt,latex-latex] (2,-2+\a/3) -- (1,-2+2*\a/3);
\end{tikzpicture}
\caption{Infrared sector of gauge theories.}
\label{Fig:IRSector}
\end{figure}

The first corner is the area of \textit{asymptotic symmetries} that has been partially studied in these notes. The second corner is the topic of \textit{soft theorems} \cite{Low:1954kd , GellMann:1954kc, Yennie:1961ad , Weinberg:1965nx , Weinberg:1995mt}. These theorems state that any $(n+1)$-particles scattering amplitude involving a massless soft particle, namely a particle with momentum $q\to 0$ (that may be a photon, a gluon or a graviton), is equal to the $n$-particles scattering amplitude without the soft particle, multiplied by the soft factor, plus corrections of order $q^0$. We have
\begin{equation}
\mathcal{M}_{n+1}(q, p_1, \ldots p_n) = S^{(0)}  \mathcal{M}_n (p_1, \ldots p_n) + \mathcal{O}(q^0) ,
\end{equation} where $S^{(0)} \sim q^{-1}$ is the soft factor whose precise form depends on the nature of the soft particle involved. Taking as soft particle a photon, gluon or graviton will respectively lead to the soft photon theorem, soft gluon theorem and soft graviton theorem. A remarkable property is that the soft factor is independent of the spin of the $n$ particles involved in the process. Furthermore, some so-called \textit{subleading soft theorems} have been established for the different types of soft particles and they provide some information about the subleading terms in $q$ \cite{Low:1958sn, Burnett:1967km , White:2011yy , Gross:1968in , Jackiw:1968zza}. They take the form
\begin{equation}
\mathcal{M}_{n+1}(q, p_1, \ldots p_n) = (S^{(0)} + S^{(1)}) \mathcal{M}_n (p_1, \ldots p_n) + \mathcal{O}(q) ,
\end{equation} where $S^{(1)} \sim  q^{0}$ is the subleading soft factor. Proposals for sub-subleading soft theorems can also be found \cite{Cachazo:2014fwa , DiVecchia:2016amo, Zlotnikov:2014sva}.

The third corner of the triangle is the topic of memory effects \cite{1974SvA , Braginsky:1986ia , Braginsky1987GravitationalwaveBW , Christodoulou:1991cr , Blanchet:1992br , Thorne:1992sdb , Favata:2010zu , Tolish:2014oda , Winicour:2014ska ,  Pate:2017vwa}. In gravity, the \textit{displacement memory effect} occurs, for example, in the passage of gravitational waves. It can be shown that this produces a permanent shift in the relative positions of a pair of inertial detectors. This shift is controlled by a field in the metric that is turned on when the gravitational wave is passing through the spacetime region of interest. Considering the Bondi gauge in asymptotically flat spacetime with boundary conditions (AF3) (equations \eqref{asymp flat 1} with \eqref{asymp flat 3}), it can be shown that
\begin{equation}
\Delta s^A \propto \Delta C_{AB} ,
\end{equation} i.e. the angular displacement $\Delta s^A$ of two inertial observers in the asymptotic region is dictated by the field $C_{AB}$. Three processes can turn on the field $C_{AB}$ and trigger an observable displacement memory effect: a variation of the Bondi mass aspect $M$ (ordinary memory effect), a burst of gravitational waves controlled by the news $N_{AB}$ (Christodoulou effect), or a burst of null matter (null memory effect) \cite{Compere:2019qed}. The analogous memory effects can also be established in electrodynamics (electromagnetic memory effect) \cite{Susskind:2015hpa , Pasterski:2015zua} and in Yang-Mills theory (color memory effect) \cite{Pate:2017vwa} where a field is turned on as a result of a burst of energy passing through the region of interest, leading to an observable phenomenon. Notice that other memory effects have been identified in gravity \cite{Pasterski:2015tva , Nichols:2018qac , Flanagan:2018yzh , Podolsky:2002sa , Podolsky:2010xh , Compere:2018ylh ,  Donnay:2018ckb , Mao:2018xcw}, including the \textit{spin memory effect} and the \textit{refraction memory effect}. 

We now briefly discuss the relation between these different topics. It has been shown that if the quantum gravity $\mathcal{S}$-matrix is invariant under the BMS symmetry \cite{Strominger:2013jfa}, then the Ward identity associated with the supertranslations is equivalent to the soft graviton theorem \cite{He:2014laa}. Furthermore, the displacement memory effect is equivalent to performing a supertranslation \cite{Strominger:2014pwa}. More precisely, the action of the supertranslation on the memory field $C_{AB}$ has the same effect as a burst of gravitational waves passing through the region of interest. This can be understood as a vacuum transition process \cite{Compere:2016jwb , Compere:2016hzt , Compere:2016gwf , Scholler:2017uni , Adjei:2019tuj}. Finally, a Fourier transform enables us to relate the soft theorem with the memory effect, which closes the triangle. This triangle controlling the infrared structure of the theory has also been constructed for other gauge theories \cite{Lysov:2014csa , Pate:2017vwa , Mao:2017tey}. Furthermore, subleading infrared triangles have been uncovered and discussed \cite{Kapec:2014opa , Lysov:2014csa , Campiglia:2016hvg , Conde:2016csj , Compere:2018ylh , Himwich:2019qmj}. In particular, the Ward identities of superrotations have been shown to be equivalent to the subleading soft graviton theorem. Furthermore, the spin memory effect and the refraction memory effect have been related to the superrotations.  

Finally, let us mention that this understanding of the infrared structure of quantum gravity is relevant to tackle the \textit{black hole information paradox} \cite{Hawking:1976ra}. Indeed, an infinite number of soft gravitons are produced in the process of black hole evaporation. Through the above correspondence, these soft gravitons are related with surface charges, called \textit{soft hairs}, that have to be taken into account in the information storage \cite{Hawking:2016msc , Hawking:2016sgy , Haco:2018ske , Haco:2019ggi , Bousso:2017dny , Mirbabayi:2016axw}.

%\section{Conclusions}

\section*{Acknowledgements} \addcontentsline{toc}{section}{Acknowledgements}

I would like to thank Glenn Barnich, Luca Ciambelli, Geoffrey Comp\`ere, Adrien Fiorucci, Yannick Herfray, K\'evin Nguyen and C\'eline Zwikel for their comments on the manuscript and discussions on related subjects. I would also like to thank the participants of the XV Modave Summer School 2019 for this stimulating edition. This work is supported by the FRIA (F.R.S.-FNRS, Belgium).

\appendix 

%\addcontentsline{toc}{section}{APPENDICES}

\section{Diffeomorphism between Bondi and Fefferman-Graham gauges}
\label{Diffeomorphism between Bondi and Fefferman-Graham gauges}

The diffeomorphism between Bondi and Fefferman-Graham gauges in asymptotically (A)dS$_4$ spacetime has been worked out explicitly in \cite{Poole:2018koa , Compere:2019bua}. In this appendix, we briefly recall how the solution space \eqref{most general sol space fg} associated with the preliminary boundary condition \eqref{as ads} in the Fefferman-Graham gauge matches with the solution space \eqref{most general sol space bondi} associated with the preliminary boundary condition \eqref{as ads Bondi} in the Bondi gauge through this diffeomorphism. The components of the three-dimensional boundary metric $g^{(0)}_{ab}$ can be expressed in terms of the functions of the Bondi gauge as
\begin{equation}
g_{tt}^{(0)} = \frac{\Lambda}{3} e^{4\beta_0} + U_0^C U^0_C, \qquad g_{tA}^{(0)} = -U_A^0, \qquad g_{AB}^{(0)} = q_{AB} .
\end{equation} Furthermore, the degrees of freedom of the stress-energy tensor $T_{ab}$ are related to degrees of freedom of the Bondi gauge as
\begin{equation}
T_{tt} \sim M, \qquad T_{tA} \sim N_A, \qquad T_{AB} \sim \mathcal{E}_{AB} .
\label{relations FG bondi}
\end{equation} The precise relations can be found in the references. Notice that the constraint $D^a_{(0)} T_{ab} = 0$ translated in terms of the functions in the Bondi gauge gives the evolution constraints with respect to the $u$ coordinate for the Bondi mass aspect $M$ and the angular momentum aspect $N_A$.

%\section{Computation of the modified Lie bracket}
%\label{Computation of the modified Lie bracket}

\section{Useful results and conventions}
\label{Useful results}

In this appendix, we establish some important frameworks and conventions. The aim of this formalism is to manipulate some local expressions, as this is convenient in field theory. We closely follow \cite{Compere:2019qed, Compere:2007az}. 

\subsection{Jet bundles}
\label{Jet bundles}

Let $M$ be the $n$-dimensional spacetime with local coordinates $x^\mu$ ($\mu = 0, \ldots , n-1$). The fields, written as $\Phi = (\phi^i)$, are supposed to be Grassmann even. The \textit{jet space} $J$ consists in the fields and the symmetrized derivatives of the fields $(\Phi, \Phi_\mu , \Phi_{\mu\nu}, \ldots)$, where $\Phi_{\mu_1 \ldots \mu_k} = \frac{\partial}{\partial x^{\mu_1}} \ldots  \frac{\partial}{\partial x^{\mu_k}} \Phi$. The \textit{symmetrized derivative} is defined as
\begin{equation}
\frac{\partial \tilde{\Phi}_{\nu_1 \ldots \nu_k}}{\partial \Phi_{\mu_1 \ldots \mu_k}}  = \delta_{(\mu_1}^{\nu_1} \ldots \delta_{\mu_k)}^{\nu_k} \delta^{\tilde{\Phi}}_\Phi .
\end{equation} In the jet space, the cotangent space at a point is generated by the variations of the fields and their derivatives at that point, namely $(\delta \Phi, \delta \Phi_{\mu}, \delta \Phi_{\mu\nu}, \ldots)$. The \textit{variational operator} is defined as
\begin{equation}
\delta = \sum_{k \ge 0} \delta \Phi_{\mu_1 \ldots  \mu_k} \frac{\partial}{\partial \Phi_{\mu_1 \ldots \mu_k}} .
\end{equation} We choose all the $\delta \Phi$, $\delta \Phi_\mu$, $\delta \Phi_{\mu\nu}, \ldots$ to be Grassmann odd, which implies that $\delta^2 = 0$. Hence, $\delta$ is seen as an exterior derivative on the jet space. 

Now, we define the \textit{jet bundle} as the fiber bundle with local trivialization $(x^\mu, \Phi,\Phi_\mu, \Phi_{\mu\nu} ,\ldots)$. Locally, the total space of the jet bundle looks like $M \times J$. A section of this fiber bundle is a map $x \to (\Phi (x), \Phi_\mu (x) , \Phi_{\mu\nu}(x), \ldots )$. The \textit{horizontal derivative} is defined as
\begin{equation}
\mathrm{d} = \mathrm{d}x^\mu \partial_\mu , \quad \text{where} \quad \partial_\mu = \frac{\partial}{\partial x^\mu} + \sum_{k \ge 0} \Phi_{\mu \nu_1 \ldots \nu_k} \frac{\partial}{\partial  \Phi_{\nu_1 \ldots \nu_k}}  .
\label{horizontal}
\end{equation} In this perspective, the variational operator can also be seen as the \textit{vertical derivative}, i.e. the derivative along the fibers. The exterior derivative on the total space can be defined as $\mathrm{d}_{Tot} = \mathrm{d} + \delta$. Notice that both $\mathrm{d}$ and $\delta$ are Grassmann odd and they anti-commute, namely
\begin{equation}
\mathrm{d} \delta = - \delta \mathrm{d} .
\end{equation} On the jet bundle, we write $\Omega^{p,q}$ the set of functions that are $p$-forms with respect to the spacetime and $q$-forms with respect to the jet space\footnote{One often refers to a \textit{variational bicomplex} structure}. 

\subsection{Some operators}

In this subsection, we introduce some additional operators used in the text and discuss their properties. 

The \textit{Euler-Lagrange} derivative of a \textit{local function} $f$, i.e. a function on the total space of the jet bundle $f = f[x,\Phi,\Phi_\mu, \Phi_{\mu\nu}, \ldots ]$, is defined as
\begin{equation}
\frac{\delta f}{\delta \Phi} = \sum_{k\ge 0} (-1)^k \partial_{\mu_1} \ldots \partial_{\mu_k} \left(\frac{\partial f}{\partial \Phi_{\mu_1 \ldots \mu_k}} \right) .
\label{Euler lagrange def}
\end{equation} This operator satisfies
\begin{equation}
\frac{\delta f}{\delta \Phi} = 0 \quad \Leftrightarrow \quad f = \partial_\mu j^\mu ,
\label{euler lagrange and total}
\end{equation} where $j^\mu$ is a local function (for a proof, see e.g. section 1.2 of \cite{Barnich:2018gdh}). 

The \textit{variation under a transformation of characteristic $Q$} (i.e. $\delta_Q \Phi = Q$) is given by
\begin{equation}
\delta_Q f = \sum_{k\ge 0} (\partial_{\mu_1} \ldots \partial_{\mu_k} Q ) \frac{\partial }{\partial \Phi_{\mu_1 \ldots \mu_k}} + (\partial_{\mu_1} \ldots \partial_{\mu_k} \delta Q ) \frac{\partial}{\partial \delta \Phi_{\mu_1 \ldots \mu_k}} .
\label{def variation}
\end{equation} The Lie bracket of characteristics is defined by $[Q_1,Q_2] = \delta_{Q_1} Q_2 - \delta_{Q_2} Q_1$ and satisfies $[\delta_{Q_1}, \delta_{Q_2}] = \delta_{[Q_1,Q_2]}$. A contracted variation of this type is Grassmann even and we have
\begin{equation}
\delta_Q  \mathrm{d} = \mathrm{d} \delta_Q , \quad \delta \delta_Q = \delta_Q \delta .
\label{commutation relation delta Q}
\end{equation} We also have the following relation between the variation under a transformation of characteristic $Q$ and the Euler-Lagrange derivative:
\begin{equation}
\delta_Q \frac{\delta f}{\delta \Phi} = \frac{\delta }{\delta \Phi} (\delta_Q f ) - \sum_{k \ge 0} (-1)^k \partial_{\mu_1} \ldots \partial_{\mu_k} \left( \frac{\partial Q}{\partial \Phi_{\mu_1\ldots\mu_k}} \frac{\delta f}{\delta \Phi} \right) .
\label{variation and euler lagrange}
\end{equation}

Let $\boldsymbol{\alpha}$ be a $(n-k)$-form on the space-time $M$. We use the notation
\begin{equation}
\boldsymbol{\alpha} = \alpha^{\mu_1 \ldots \mu_{k}} (\mathrm{d}^{n-k}x)_{\mu_1 \ldots \mu_k} ,
\end{equation} where
\begin{equation}
(\mathrm{d}^{n-k}x)_{\mu_1 \ldots \mu_k} = \frac{1}{k!(n-k)!} \epsilon_{\mu_1 \ldots \mu_k \nu_1 \ldots \nu_{n-k}} \mathrm{d}x^{\nu_1} \wedge \ldots \wedge \mathrm{d}x^{\nu_{n-k}}
\end{equation} and where $\epsilon_{\mu_1 \ldots \mu_n}$ is completely antisymmetric and $\epsilon_{01\ldots n-1} = 1$. We can check that
\begin{equation}
\mathrm{d} \boldsymbol{\alpha} = \partial_\sigma {\alpha}^{[\mu_1 \ldots \mu_{k-1} \sigma]} (\mathrm{d}^{n-k+1}x)_{\mu_1 \ldots \mu_{k-1}} .
\end{equation} The \textit{interior product of a spacetime form} with respect to a vector field $\xi$ is defined as
\begin{equation}
\iota_\xi = \xi^\mu \frac{\partial}{\partial \mathrm{d}x^\mu} .
\end{equation} Notice that we can also define the \textit{interior product of a jet space form} with respect to a characteristic $Q$ as
\begin{equation}
i_Q = \sum_{k\ge 0} ( \partial_{\mu_1} \ldots \partial_{\mu_k} Q ) \frac{\partial}{\partial \delta \Phi_{\mu_1 \ldots \mu_k}} .
\end{equation} It satisfies
\begin{equation}
i_Q \delta + \delta i_Q = \delta_Q, \quad i_{Q_1} \delta_{Q_2} - \delta_{Q_2} i_{Q_1} = i_{[Q_1,Q_2]} .
\end{equation}

The \textit{homotopy operator} $I_{\delta \Phi}^p : \Omega^{p,q} \mapsto \Omega^{p-1, q+ 1}$ is defined as
\begin{equation}
I_{\delta \Phi}^p \boldsymbol{\alpha} = \sum_{k\ge 0} \frac{k + 1}{n-p+ k +1} \partial_{\mu_1} \ldots \partial_{\mu_k} \left( \delta \Phi \frac{\delta}{\delta \Phi_{\mu_1\ldots\mu_k \nu}} \frac{\partial \boldsymbol{\alpha}}{\partial \mathrm{d}x^\nu} \right)
\label{homotopy operator}
\end{equation} for $\boldsymbol{\alpha} \in \Omega^{p,q}$. This operator satisfies the following relations
\begin{align}
\delta &= \delta \Phi \frac{\delta}{\delta \Phi} - \mathrm{d} I^n_{\delta \Phi} \quad \text{when acting on spacetime $n$-forms} , \\
\delta &= I^{p+1}_{\delta \Phi} \mathrm{d} - \mathrm{d} I_{\delta \Phi}^p \quad \text{when acting on spacetime $p$-forms ($p<n$)} .
\label{commutation homotopy op}
\end{align} Furthermore,
\begin{equation}
\delta I^p_{\delta \Phi} = I^p_{\delta \Phi} \delta .
\end{equation} Notice that the homotopy operator is used to prove the algebraic Poincar\'e lemma \eqref{Poincare lemma}.

Similarly, the \textit{homotopy operator with respect to gauge parameters $F$} is defined as $I_F^p : \Omega^{p,q} \mapsto \Omega^{p-1,q}$, where
\begin{equation}
I^p_F \boldsymbol{\alpha} = \sum_{k\ge 0} \frac{k + 1}{n-p+ k +1} \partial_{\mu_1} \ldots \partial_{\mu_k} \left(F \frac{\delta}{\delta F_{\mu_1\ldots\mu_k \nu}} \frac{\partial \boldsymbol{\alpha}}{\partial \mathrm{d} x^\nu} \right) .
\end{equation} It satisfies
\begin{equation}
I^{p+1}_F \mathrm{d} + \mathrm{d} I^p_F  = 1 .
\end{equation}

%%\bibliographystyle{unsrt}
%\bibliographystyle{utphys}
%\bibliography{BiblioModave}

\addcontentsline{toc}{section}{References}

\providecommand{\href}[2]{#2}\begingroup\raggedright\endgroup

\end{document}